\documentclass[preprint,12pt]{elsarticle}

\usepackage{amssymb}
\usepackage{float}
\usepackage{latexsym}
\usepackage[section]{placeins}
\usepackage{url}
\usepackage{xcolor}
\usepackage{hyperref}
\usepackage{amsmath}
\usepackage{gensymb}
\usepackage{breqn}
\usepackage{booktabs}
\usepackage{threeparttable} 
\usepackage{algorithm}
\usepackage{algpseudocode}
\usepackage{xfrac}
\usepackage{pifont}
\usepackage{amsmath}
\usepackage{multirow}
\usepackage{hyperref}

\journal{Medical Image Analysis}

\begin{document}

\begin{frontmatter}



\title{Robust Simultaneous Multislice MRI Reconstruction Using Slice-Wise Learned Generative Diffusion Priors}

\author[label1]{Shoujin Huang}
\author[label2]{Guanxiong Luo}
\author[label3]{Yunlin Zhao}
\author[label3]{Yilong Liu}
\author[label1]{Yuwan Wang}
\author[label1]{Kexin Yang}
\author[label5]{Jingzhe Liu}
\author[label6]{Hua Guo}
\author[label7,label8]{Min Wang}
\author[label4]{Lingyan Zhang\corref{cor1}}
\ead{18819818005@163.com}
\author[label1]{Mengye Lyu\corref{cor1}}
\ead{lvmengye@sztu.edu.cn}
\cortext[cor1]{Corresponding authors}
\affiliation[label1]{%
   organization={College of Health Science and Environmental Engineering, Shenzhen Technology University},
   city={Shenzhen},
   country={China}
}

\affiliation[label2]{%
   organization={University Medical Center Göttingen},
   city={Göttingen},
   country={Germany}
}

\affiliation[label3]{%
   organization={Guangdong-Hongkong-Macau CNS Regeneration Institute, Key Laboratory of CNS Regeneration (Jinan University)-Ministry of Education, Jinan University},
   city={Guangzhou},
   country={China}
}

\affiliation[label5]{%
   organization={Department of Radiology, The First Hospital of Tsinghua University},
   city={Beijing},
   country={China}
}

\affiliation[label6]{%
   organization={Center for Biomedical Imaging Research, Department of Biomedical Engineering, School of Medicine, Tsinghua University},
   city={Beijing},
   country={China}
}

\affiliation[label7]{%
   organization={Key Laboratory for Biomedical Engineering of Ministry of Education, College of Biomedical Engineering and Instrument Science, Zhejiang University},
   city={Hangzhou},
   country={China}
}

\affiliation[label8]{%
   organization={Department of Endocrinology and Metabolism, Sir Run Run Shaw Hospital, Zhejiang University School of Medicine},
   city={Hangzhou},
   country={China}
}

\affiliation[label4]{%
   organization={Lab of Molecular Imaging and Medical Intelligence, Department of Radiology, Longgang Central Hospital of Shenzhen (Shenzhen Clinical Medical College, Guangzhou University of Chinese Medicine; Longgang Clinical Institute of Shantou University Medical College)},
   city={Shenzhen},
   country={China}
}



\begin{abstract}
Simultaneous multislice (SMS) imaging is a powerful technique for accelerating magnetic resonance imaging (MRI) acquisitions. However, SMS reconstruction remains challenging due to complex signal interactions between and within the excited slices. In this study, we introduce ROGER, a robust SMS MRI reconstruction method based on deep generative priors. Utilizing denoising diffusion probabilistic models (DDPM), ROGER begins with Gaussian noise and gradually recovers individual slices through reverse diffusion iterations while enforcing data consistency from measured k-space data within the readout concatenation framework. The posterior sampling procedure is designed such that the DDPM training can be performed on single-slice images without requiring modifications for SMS tasks. Additionally, our method incorporates a low-frequency enhancement (LFE) module to address the practical issue that SMS-accelerated fast spin echo (FSE) and echo planar imaging (EPI) sequences cannot easily embed fully-sampled autocalibration signals. Extensive experiments on both retrospectively and prospectively accelerated datasets demonstrate that ROGER consistently outperforms existing methods, enhancing both anatomical and functional imaging with strong out-of-distribution generalization. The source code and sample data for ROGER are available at \url{https://github.com/Solor-pikachu/ROGER}. 
\end{abstract}

\begin{graphicalabstract}
\centering
    \includegraphics[width=\textwidth]{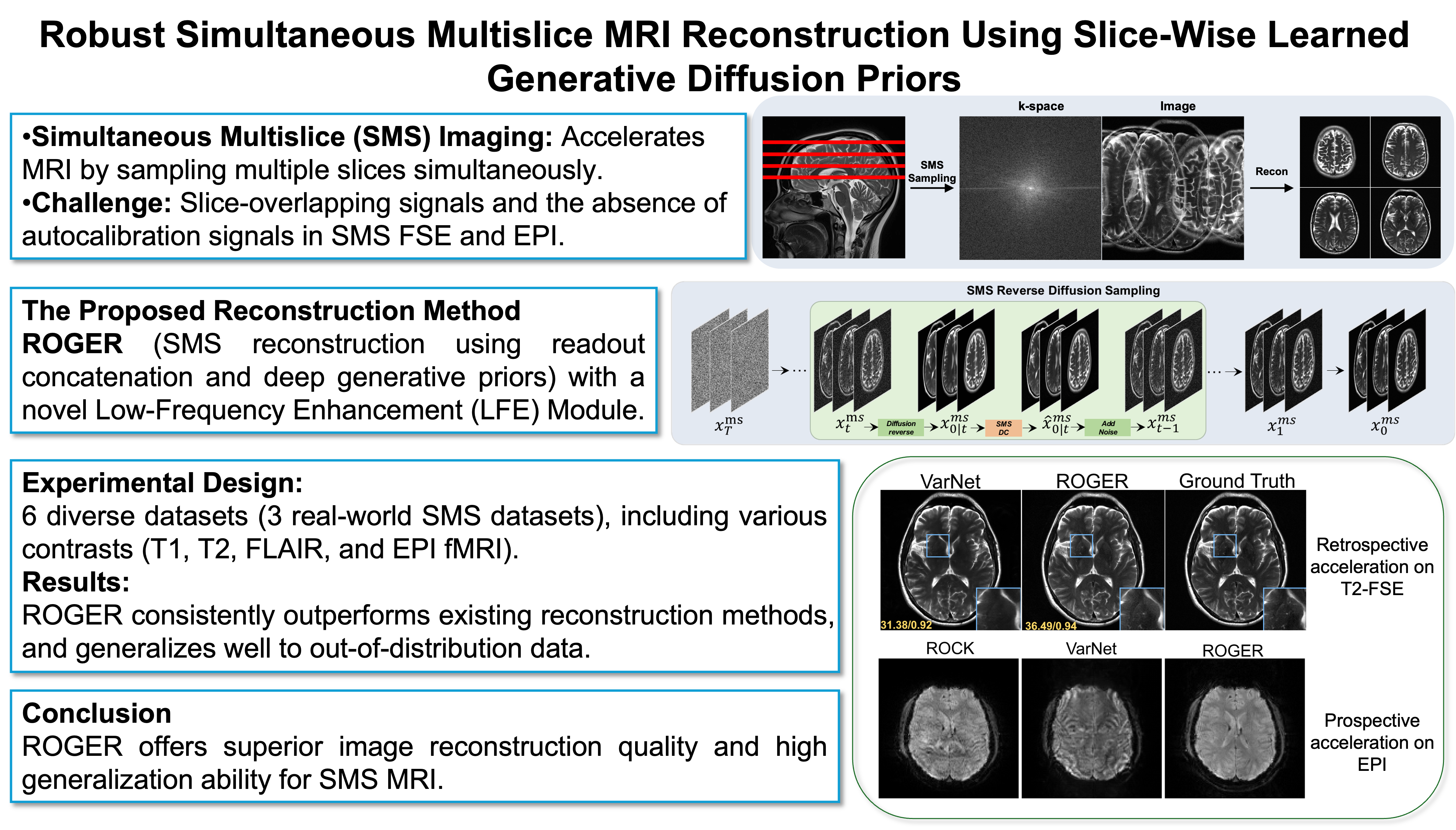}
    \label{fig:graphical_abstract}
\end{graphicalabstract}

\begin{highlights}

\item A novel SMS MRI reconstruction method with generative priors via diffusion model.
\item A Low-Frequency Enhancement module to stabilize reconstruction for real-world data.
\item Comprehensive validation on both anatomical and functional MRI using 6 datasets.
\item Superior reconstruction quality with strong out-of-distribution generalization.

\end{highlights}


\end{frontmatter}



\section{Introduction}
Accelerating magnetic resonance imaging (MRI) is important for capturing subtle spatial/temporal information, improving patient throughput, and minimizing motion artifacts. Simultaneous multislice (SMS) imaging\citep{breuer2005controlled,moeller2014ro,setsompop2012blipped,barth2016simultaneous} addresses this by using multiband (MB) radio-frequency pulses to acquire multiple slices simultaneously, effectively reducing scan time and/or improving slice coverage. Unlike in-plane acceleration, which suffers from intrinsic signal-to-noise ratio (SNR) loss due to k-space undersampling, SMS acquisitions benefit from improved SNR efficiency due to Fourier averaging\citep{barth2016simultaneous}, and have been widely used for anatomical, diffusion-weighted, and functional MRI, including the Human Connectome Project (HCP)\citep{moeller2010multiband,HCPvan2012human,HCPharms2018extending}.

Despite its advantages, SMS MRI presents considerable reconstruction challenges. The simultaneous acquisition of multiple slices results in inter-slice signal interactions and potential artifacts. Traditional SMS reconstruction methods\citep{breuer2005controlled,blaimer2006accelerated,moeller2010multiband,setsompop2012blipped,cauley2014interslice} are adapted from classical parallel imaging techniques, often suffering from noise amplification and residual aliasing artifacts. Improved iterative approaches\citep{SMS-NLINV,demirel2021improved,park2017sms,LIM2022102621,demirel2023improved} introduce various regularizations to stabilize the reconstruction but still struggle with ill-conditioning at high acceleration settings.

\begin{figure*}[t]
\centering
\includegraphics[width=1.0\linewidth]{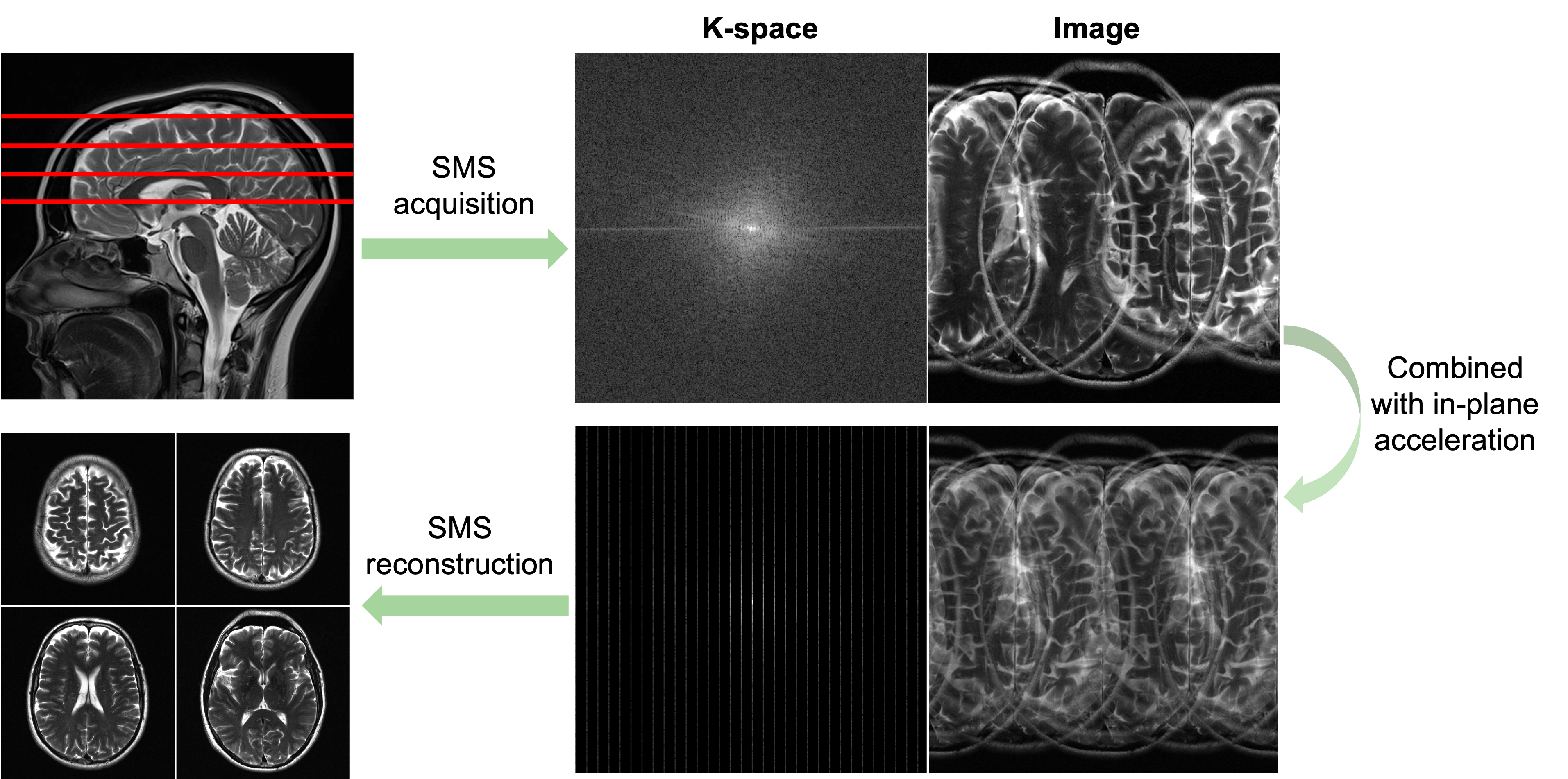}
\caption{Illustration of the simultaneous multislice (SMS) MRI acquisition and reconstruction process. A case with four-fold SMS acceleration and two-fold in-plane acceleration is plotted as an example. The acquisition involves multiple non-adjacent slices with CAIPI (controlled aliasing in parallel imaging) shift patterns and additional in-plane acceleration, resulting in sparse k-space and complex aliasing artifacts. SMS reconstruction is difficult due to the strong aliasing artifacts and the absence of fully-sampled autocalibration signals in many SMS-accelerated sequences.}
\label{fig:sms_illu}
\end{figure*}

Recent advancements in deep learning have shown promise for improving MRI reconstruction quality\citep{Liang2020Review}. Supervised learning methods\citep{aggarwal2018modl,hammernik2018varnet,sriram2020e2evarnet} have been successfully demonstrated on in-plane accelerated MRI, leveraging large datasets to learn the reconstruction mappings. These models may also be adapted to SMS reconstruction by training on SMS-accelerated and fully sampled k-space pairs\citep{LE2021178unetSMS} or in a self-supervised manner\citep{demirel202120}. However, supervised deep learning methods may not generalize well to unseen data, such as data with different acquisition parameters, aliasing patterns, and coil sensitivity distributions.

Recently, the use of generative models has emerged as a potentially more robust approach for MRI reconstruction\citep{song2021solving,jalal2021robust,chung2022score,luo2023bayesian,gungor2023adaptive,korkmaz2023self,cao2024high,luo2024autoregressive}. Generative models can learn data distributions as priors and solve various inverse problems\citep{song2021solving}. However, the application of generative models to SMS reconstruction remains unexplored and presents several challenges for the following reasons.

Firstly, the forward model of SMS imaging differs from conventional k-space subsampling. This process includes encoding multiple 2D slices by different coil sensitivity maps, phase cycling for Controlled Aliasing in Parallel Imaging Results in Higher Acceleration (CAIPI) shifts\citep{breuer2005controlled}, and the summation of signals into a single 2D matrix. This complexity, combined with the variations in MB factors (number of simultaneously acquired slices), CAIPI shift patterns, and additional in-plane acceleration factors, can result in a wide range of aliasing patterns and very sparse k-space, as illustrated by a typical SMS dataset in Fig.~\ref{fig:sms_illu}. Moreover, for real-world SMS-accelerated data, it remains unclear whether system imperfections such as slice excitation side-lobes\citep{park2016new,SMSspiral} and phase errors between echoes\citep{koopmans2017two,hoge2018dual,lyu2018improved} may interfere with diffusion models, potentially amplifying artifacts during reverse diffusion. These imperfections are often hardware-specific and difficult to simulate, making real SMS data essential for evaluating diffusion model-based SMS reconstruction.

Secondly, SMS acceleration is desirable across diverse MRI applications, including anatomical imaging and functional/diffusion-weighted imaging\citep{fMRI7TeslaReview,ye2024abdominal, DWI_SMS_MIA, fMRI_SMS_MIA}. These applications may have very different imaging parameters and tissue contrasts. For instance, anatomical imaging by T2-weighted fast spin echo (FSE) typically uses large matrix sizes over 300, while functional echo planar imaging (EPI) often uses smaller matrix sizes around 100. Ensuring high-quality reconstructions across these scenarios requires highly robust reconstruction methods.

Thirdly, integrating autocalibration signals (ACS, i.e., fully sampled central k-space) within SMS-accelerated scans is difficult\citep{SMSLORAKS,barth2016simultaneous}, leading to very sparse k-space with limited low-frequency information in real-world SMS data. While some solutions have been proposed for gradient echo sequences\citep{ferrazzi2019autocalibrated,rapacchi2019simultaneous,ferrazzi2020autocalibrated,zou2023myocardial,gaspar2023open}, this problem remains cumbersome in Cartesian FSE\citep{fritz2017simultaneous} and EPI\citep{setsompop2012blipped,cauley2014interslice,demirel202120} sequences, which are two of the most frequently used MRI sequences in clinical and preclinical research. In FSE, integrating ACS complicates phase encoding and requires careful optimization to address signal discontinuities between the echoes. In EPI, uniform k-space sampling is preferred to reduce geometric distortions because of the rapid T2* decay\citep{zhang2023calibrationless}. Therefore, a separate single-band scan (external calibration) is typically used to provide coil sensitivity maps for SMS reconstruction\citep{barth2016simultaneous}, as done in numerous SMS methodological and application studies, as well as in vendor-provided SMS sequences\citep{moeller2010multiband,hoge2018dual,lyu2018robust,liu2019pec,muftuler2020optimization}.
However, such calibration scans may not carry the same tissue contrasts and image phases as the SMS scan, such that they cannot be directly merged into the SMS data to provide low-frequency information, rendering reconstruction challenging.

To address these challenges, we propose a novel SMS MRI reconstruction method based on denoising diffusion probabilistic models (DDPM) and validate it through extensive experiments in this study. The contributions are summarized as follows:
\begin{itemize}
\item We present a new reconstruction method that employs deep generative priors to separate the highly aliased slices in SMS MRI. We name this method ROGER (SMS reconstruction using ReadOut concatenation and deep GEneRative priors). 

\item We propose a Low-Frequency Enhancement (LFE) module that stabilizes the reverse diffusion process, particularly benefiting widely used FSE and EPI sequences that cannot easily integrate ACS.

\item Compared with existing methods, our method demonstrates substantially improved image SNR, fewer artifacts, and strong generalization ability across six datasets under various settings.
\end{itemize}

\section{BACKGROUND}
\subsection{Related Work and Problem Formulation}
SMS MRI reconstruction can be approached using several different frameworks\citep{SMSbook,larkman2001use,setsompop2012blipped,blaimer2006accelerated,moeller2014ro,Zhu3DSMS} for handling the complexities of SMS encoding and signal separation.

Classical slice-SENSE method\citep{larkman2001use,breuer2005controlled} utilizes known coil sensitivity profiles to separate signals from simultaneously acquired slices in pure image domain. GRAPPA-based methods\citep{griswold2002generalized} offer another approach to SMS reconstruction based on k-space interpolation. While SENSE-GRAPPA\citep{blaimer2006accelerated} extends the field of view (FOV) along the phase encoding direction and applies traditional GRAPPA methods to solve for aliased signals, slice-GRAPPA\citep{setsompop2012blipped} reconstructs the data directly into separated slices without extending FOV. An improved implementation of slice-GRAPPA is split slice-GRAPPA (SPSG)\citep{cauley2014interslice}, which trains kernels not only to reconstruct the target slice, but also to suppress erroneous mappings to other slices. This approach mitigates the slice leakage issue and improves temporal signal-to-noise ratio (tSNR).

SMS reconstruction can also be reformulated using the readout concatenation framework\citep{moeller2014ro,koopmans2017two,lyu2018robust,liu2019pec,demirel2021improved}, which transforms SMS encoding as a one-dimensional in-plane acceleration along the readout direction. In this approach, the slices to be reconstructed are viewed as spatially concatenated, forming a single 2D image with the FOV extended MB times along the readout direction. Consequently, SMS acceleration can be characterized as a uniform k-space subsampling in this extended readout direction, with optional in-plane undersampling incorporated in the phase encoding dimension. Thus, the forward model for ROC based SMS reconstruction is given by:
\begin{equation}
\label{eq:2}
\mathbf{y} = \mathcal{A} \mathcal{R} {\bar{\mathbf{x}}^{ms}} + \mathbf{n}
\end{equation}
, where $\bar{\mathbf{x}}^{ms} := \{\bar{\mathbf{x}}^{1},\bar{\mathbf{x}}^{2},\bar{\mathbf{x}}^{3},...,\bar{\mathbf{x}}^{MB}\}$ denotes the slice images to be reconstructed, $\mathbf{n}$ is complex-valued Gaussian noise, $\mathcal{R}$ represents the combined data reorder operations of $\mathcal{ROC}$ (readout concatenation) and $\mathcal{CS}$ (CAIPI shift), and $\mathcal{A}$ is the SENSE encoding operator. $\mathcal{A}$ can be decomposed into $\mathcal{P} \cdot \mathcal{F} \cdot \mathcal{S}$ where $\mathcal{F}$ is two-dimensional Fourier transform, $\mathcal{S}$ coil sensitivity maps, and $\mathcal{P}$ k-space subsampling. We adopt this ROC framework for applying SMS data consistency terms in this study.

\subsection{Denoising Diffusion Probabilistic Models}
Diffusion models are probabilistic generative models that express image generation as a temporal Markov process. DDPM defines a $T$-step forward and reverse diffusion process\citep{dhariwal2021diffusion}. The forward process adds random Gaussian noise to image, while the reverse process constructs desired data samples from the Gaussian noise. 
\begin{figure*}[t]
\centering
\includegraphics[width=1.0\linewidth]{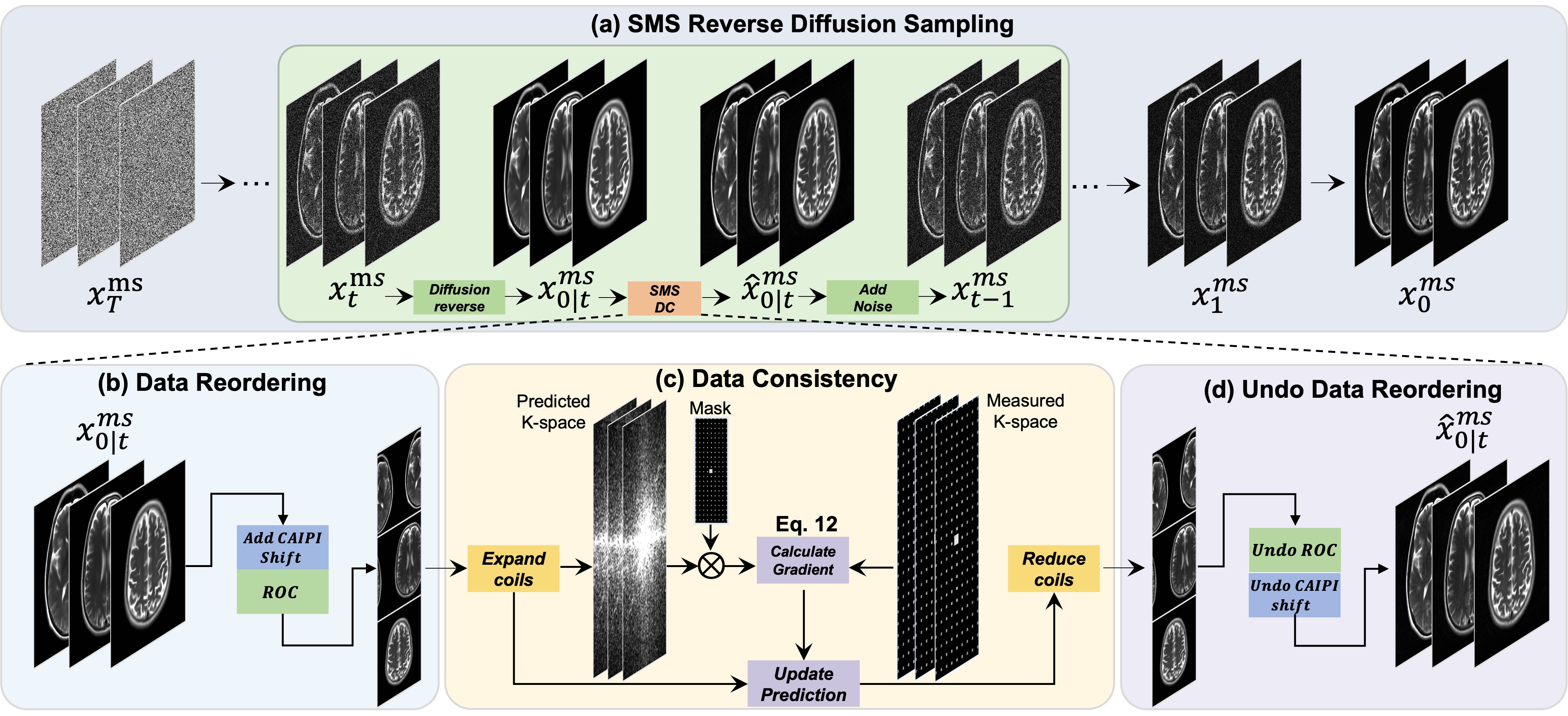}
\caption{Schematic illustration of the proposed ROGER method. (a) The overall reconstruction procedure using reverse diffusion sampling: at each reverse diffusion sampling timestep $t$, the image estimate $\mathbf{x}_{0|t}^{ms}$ is refined from $\mathbf{x}_{t}^{ms}$ using pre-trained denoising model via Eq.~\ref{eq:ROGER1}, followed by an adjustment with the SMS data consistency (DC) term from Eq.~\ref{eq:13}. Noise is then added to $\mathbf{\hat{x}}_{0|t}^{ms}$ to compute $\mathbf{x}_{t-1}^{ms}$ for the next sampling step as per Eq.~\ref{eq:ROGER2}. The SMS DC process includes (b) Data Reordering $\mathcal{R}$, which applies CAIPI shift and readout concatenation (ROC) to $\mathbf{x}_{0|t}^{ms}$, (c) Data Consistency, which updates $\mathcal{R}\mathbf{x}_{0|t}^{ms}$ using the gradient of the DC term, and (d) Undo Data Reordering $\mathcal{R}^{H}$, which reverses the reordering process to derive $\mathbf{\hat{x}}_{0|t}^{ms}$, facilitating continuation of the reverse diffusion sampling.}
\label{fig:method}
\end{figure*}
\subsubsection{Forward Diffusion and Prior Training}
The forward process yields the present state $\mathbf{x}_{t}$ from the previous state $\mathbf{x}_{t-1}$. At step \( t \), the relationship between \( t \) and \( t-1 \), along with the conditional distribution \( q(x_{t} \mid x_{t-1}) \), is specified as follows:
\begin{equation}
q(\mathbf{x}_{t}|\mathbf{x}_{t-1}) = \mathcal{CN}(\mathbf{x}_{t};\sqrt{1-\beta_{t}}\mathbf{x}_{t-1}, \beta_{t}\mathbf{I})
\end{equation}
\begin{equation}
x_{t}=\sqrt{1-\beta_{t}}\mathbf{x}_{t-1} + \sqrt{\beta_{t}}\mathbf{z}, \mathbf{z} \sim \mathcal{CN}(0, \mathbf{I})~,
\end{equation}
where $\beta_{t}$ follows a pre-defined schedule $\{\beta_{0},\beta_{1}, ... , \beta_{T}\}$ which is an increasing sequence of t. \(\mathbf{x}_{t}\) converges to isotropic Gaussian noise after a large number of forward steps. Following re-parameterization method\citep{ho2020denoising}, the two equations become:
\begin{equation}
\label{eq:5}
q(\mathbf{x}_{t}|\mathbf{x}_{0}) = \mathcal{CN}(\mathbf{x}_{t};\sqrt{\overline{a}_{t}}\mathbf{x}_{0}, (1-\overline{a}_{t})\mathbf{I})
\end{equation}
\begin{equation}
\label{eq:6}
x_{t}=\sqrt{\overline{a}_{t}}\mathbf{x}_{0} + \sqrt{1-\overline{a}_{t}}\mathbf{z}, \mathbf{z} \sim \mathcal{CN}(0, \mathbf{I})~,
\end{equation}
with $\overline{a}_{t}:=\prod \limits_{s=1}^{t} {a}_{s}$, ${a}_{t}:=1-\beta_{t}$.

To obtain the generative prior, DDPM trains a neural network $\epsilon^{t}_{\theta}$ to predict the noise $\mathbf{z}$ for each time-step $t$, and $\epsilon^{t}_{\theta}$ is then used in the reverse diffusion process. For every training step, DDPM randomly picks a clean image $\mathbf{x}_{0}$ from the high-quality dataset and samples a noise $\mathbf{z} \sim \mathcal{CN}(0, \mathbf{I})$, then picks a random time-step t and updates the network parameters $\theta$ in $\epsilon^{t}_{\theta}$ to minimize the following expectation:
\begin{equation}
\overline{}min \mathbb{E}_{x_{t} \sim q(\mathbf{x}_{t}|\mathbf{x}_{0}), \mathbf{x}_{0} \sim p_{data}(\mathbf{x}_{0}),\mathbf{z} \sim \mathcal{CN}(0, \mathbf{I})} {\left|\left| \epsilon^{t}_{\theta}(\mathbf{x}_{t}-\mathbf{z}) \right|\right|^{2}_{2}} 
\end{equation}

\subsubsection{Reverse Diffusion and Image Generation}
The reverse process aims at yielding the previous state $\mathbf{x}_{t-1}$ from $\mathbf{x}_{t}$ using the posterior distribution $p(\mathbf{x}_{t-1}|\mathbf{x}_{t}, \mathbf{x}_{0})$, which can be derived from the Bayes theorem from Eqs.~\ref{eq:5} and ~\ref{eq:6}:
\begin{align}
p(\mathbf{x}_{t-1}|\mathbf{x}_{t}, \mathbf{x}_{0}) &= q(\mathbf{x}_{t}|\mathbf{x}_{t-1}) \frac{q(\mathbf{x}_{t-1}|\mathbf{x}_{0})}{q(\mathbf{x}_{t}|\mathbf{x}_{0})} \nonumber \\
&= \mathcal{CN}(\mathbf{x}_{t-1};\mu_{t}(\mathbf{x}_{t},\mathbf{x}_{0}),\sigma^{2}_{t}\mathbf{I})
\end{align}
with the closed forms of mean $\mu_{t}(\mathbf{x}_{t},\mathbf{x}_{0})= \frac{1}{{\sqrt{a_{t}}}}({\mathbf{x}_{t}-\frac{1-a_{t}}{\sqrt{1-\overline{a}_{t}}}\epsilon^{t}_{\theta}(\mathbf{x}_{t})})$, and variance $\sigma^{2}_{t}=\frac{1-\overline{a}_{t-1}}{1-\overline{a}_{t}}\beta_{t}$. Hence, we have
\begin{equation}
\label{eq:DDPM reverse}
\mathbf{x}_{t-1} = \frac{1}{{\sqrt{a_{t}}}}({\mathbf{x}_{t}-\frac{1-a_{t}}{\sqrt{1-\overline{a}_{t}}}\epsilon^{t}_{\theta}(\mathbf{x}_{t})}) + \frac{1-\overline{a}_{t-1}}{1-\overline{a}_{t}}\beta_{t}\mathbf{z}
\end{equation}
We adopt denoising diffusion implicit model (DDIM) as the reverse sampling method, as DDIM is known to be suitable for point estimation of the posterior without requiring multiple runs\citep{song2020denoising}. Following previous work\citep{wang2022zero,DDS}, the reverse diffusion can be described as:
\begin{equation}
\label{eq:ddpm reverse1}
\mathbf{x}_{0|t}=\frac{1}{{\sqrt{\overline{a}_{t}}}}({\mathbf{x}_{t}-\sqrt{1-\overline{a}_{t}}\epsilon^{t}_{\theta}(\mathbf{x}_{t})})
\end{equation}
\begin{equation}
\label{eq:ddpm reverse2}
\mathbf{x}_{t-1} = \sqrt{\overline{a}_{t-1}}\mathbf{x}_{0|t} + \sqrt{1-\overline{a}_{t-1}}(\sqrt{1-\eta^2}\epsilon^{t}_{\theta}(\mathbf{x}_{t}) + \eta \mathbf{z})
\end{equation}
where $\mathbf{x}_{0|t}$ is the denoised estimate $\mathbf{x}_{t}$ at time-step $t$, and $\eta$ is a hyperparameter that controls the randomness of the sampling process\citep{song2020denoising,wang2022zero}. By iteratively sampling $\mathbf{x}_{t-1}$ from $p(\mathbf{x}_{t-1}|\mathbf{x}_{t},\mathbf{x}_{0})$, DDPM can yield clean images $\mathbf{x}_{0} \sim q(\mathbf{x})$ from
 random Gaussian noises $\mathbf{x}_{T} \sim \mathcal{CN}(0, \mathbf{I})$, where $q(\mathbf{x})$ is the approximation of the distribution of training data.

 For MRI reconstruction and many image restoration/enhancement tasks alike, it is necessary to equip the above iterations with data consistency (fidelity) guidance to achieve conditional image generation\citep{song2021solving,jalal2021robust,chung2022score,luo2023bayesian,gungor2023adaptive,korkmaz2023self,cao2024high,huang2024noise,garber2024image,wang2022zero,kawar2022denoising,song2023pseudoinverse}. As such, the reconstructed image is considered a sample drawn from the posterior distribution conditioned on the measured data. This posterior distribution consists of two components: 1) the likelihood term, which models data consistency with the measured data, and 2) the diffusion model-based generative prior, which models the distribution of images based on prior knowledge learned from the training dataset\citep{luo2023bayesian,luo2023generative}. Following this paradigm, our proposed SMS reconstruction method (ROGER) is described in the following section.

\section{METHODOLOGY}
\subsection{SMS Reverse Diffusion Sampling}\label{reverse_sampling}
As illustrated in Fig.~\ref{fig:method}, the ROGER reconstruction method employs DDPM to provide slice-wise learned probability distributions, and the data consistency term is applied using the ROC framework.

Our goal is to sample a proper $\bar{\mathbf{x}}^{ms} := \{\bar{\mathbf{x}}^{1},\bar{\mathbf{x}}^{2},\bar{\mathbf{x}}^{3},...,\bar{\mathbf{x}}^{MB}\}$ from the learned probability distributions conditioned on the SMS MRI measurements $\mathbf{y}$. To use Eq.~\ref{eq:ddpm reverse1}, we initialize the $\mathbf{x}^{ms}_{t}$ from random Gaussian noise $\mathbf{z} \sim \mathcal{CN}(0,\mathbf{I})$, and the (unconditional) reverse diffusion process is 
\begin{equation}
\label{eq:ROGER1}
\mathbf{x}^{ms}_{0|t}=\frac{1}{{\sqrt{\overline{a}_{t}}}}({\mathbf{x}^{ms}_{t}-\sqrt{1-\overline{a}_{t}}\epsilon^{t}_{\theta}(\mathbf{x}^{ms}_{t})})
\end{equation}

To incorporate guidance from measured data as described by the SMS forward model in Eq.~\ref{eq:2}, the gradient of the data consistency term is used\citep{song2021solving,jalal2021robust,chung2022score,luo2023bayesian,song2023pseudoinverse}. Specifically, we enforce that $\mathbf{x}^{ms}_{0|t}$ satisfies the data consistency constraint $\mathcal{(AR)}^{H}\mathcal{AR}\mathbf{x}^{ms}_{0|t} = \mathcal{(AR)}^{H}\mathbf{y}$ by gradient descent, resulting in
\begin{equation}
\label{eq:13}
\mathbf{\hat{x}}^{ms}_{0|t} = \mathbf{x}^{ms}_{0|t} - \lambda\mathcal{(AR)}^{H}(\mathcal{AR}\mathbf{x}^{ms}_{0|t} - \mathbf{y})
\end{equation}
where $\mathcal{(AR)}^{H}$ denotes the adjoint-inverse of $\mathcal{AR}$ and $\lambda$ is the guidance scaling factor. This hyperparameter $\lambda$ is fixed to be 2 in this study. Then, based on Eq.~\ref{eq:ddpm reverse2}, we yield $\mathbf{x}^{ms}_{t-1}$ by sampling from $p(\mathbf{x}^{ms}_{t-1}|\mathbf{x}^{ms}_{t},\mathbf{\hat{x}}^{ms}_{0|t})$:
\begin{equation}
\label{eq:ROGER2}
\mathbf{x}^{ms}_{t-1} = \sqrt{\overline{a}_{t-1}}\mathbf{\hat{x}}^{ms}_{0|t} + \sqrt{1-\overline{a}_{t-1}}(\sqrt{1-\eta^2}\epsilon^{t}_{\theta}(\mathbf{x}^{ms}_{t}) + \eta \mathbf{z}), \mathbf{z} \sim \mathcal{CN}(0,\mathbf{I})
\end{equation}
By applying Eqs.~\ref{eq:ROGER1},~\ref{eq:13} and~\ref{eq:ROGER2} sequentially and iteratively, we can finally yield the desired sample $\mathbf{x}^{ms}_{0}$ at time-step $0$, and this sample is the aliasing-free reconstructed images without slice overlapping.

\subsection{Low-Frequency Enhancement Module}

\begin{figure*}[t]
\centering
\includegraphics[width=0.8\linewidth]{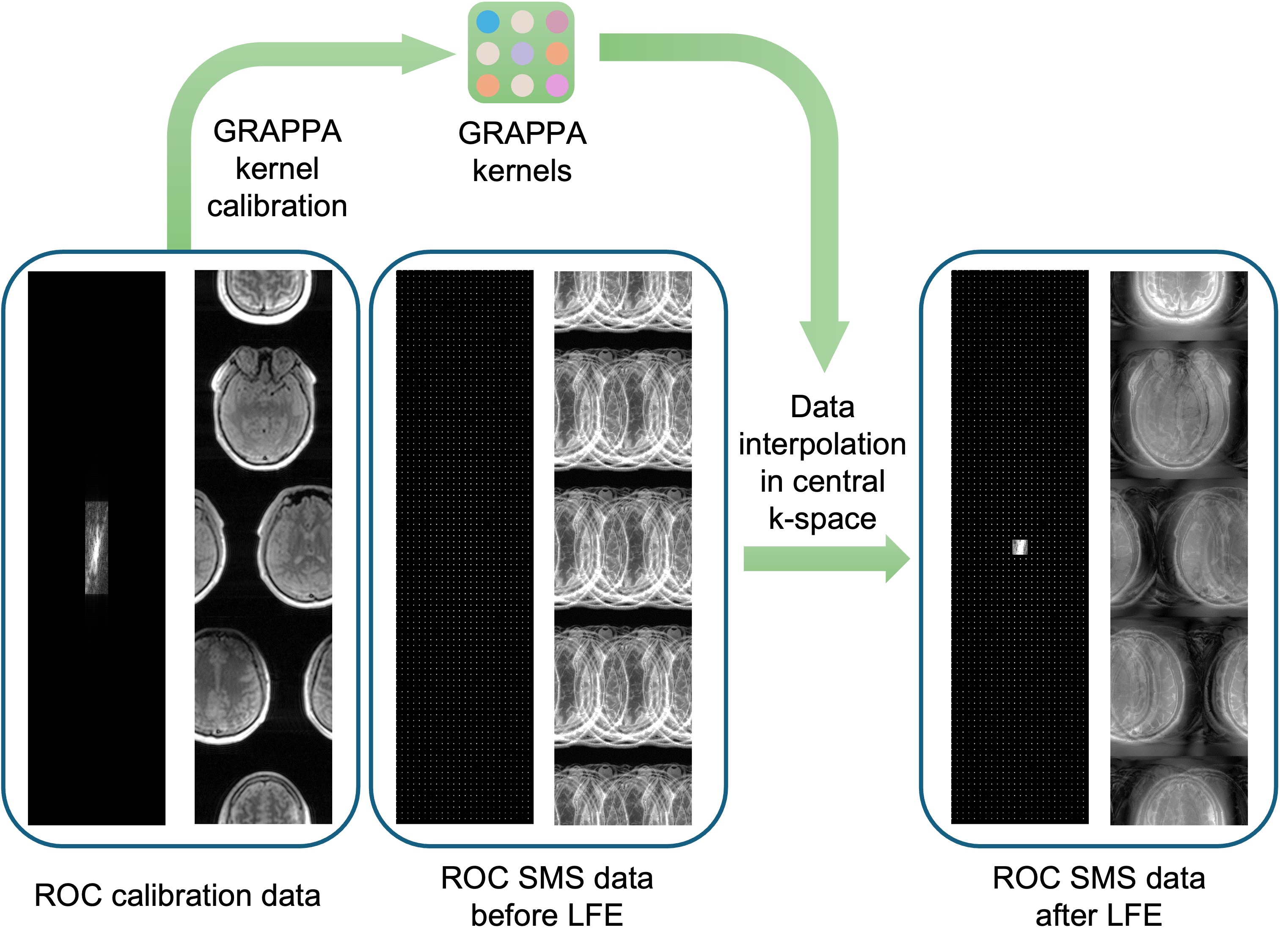}
\caption{Procedure of the proposed Low-Frequency Enhancement (LFE) module. A case with four-fold SMS acceleration and two-fold in-plane acceleration is plotted as an example. GRAPPA kernels are trained from the readout concatenated (ROC) calibration data, and applied to the ROC SMS data to interpolate the missing data points in the central k-space, thus partially recovering the low-frequency information. A set of SMS FSE data and its vendor-provided calibration data are displayed for visualization with the k-space plotted on the left and images on the right. Note that the SMS data are T2-weighted yet the calibration data are close to proton-density weighted. The coil dimension is not plotted for simplicity.}
\label{fig:LFE_illu}
\end{figure*}

In practice, SMS-accelerated FSE and EPI sequences cannot easily embed autocalibration signals (ACS). The absence of low-frequency renders reconstruction challenging as the information for object shape and tissue contrasts is not available. To address this, we propose a straightforward yet effective low-frequency enhancement (LFE) method, utilizing GRAPPA for interpolation in the central k-space of the ROC data. The procedure of this LFE module is schematically illustrated in Fig.~\ref{fig:LFE_illu}. The first step is to estimate the GRAPPA kernels from a calibration scan using linear least-squares fitting. This process is described mathematically as follows:
\begin{equation}
\label{eq:kernel for grappa}
kernel_{\theta} \xleftarrow{\text{fitting}} \mathcal{R} (\mathbf{y}_{calib})
\end{equation}
where $\mathbf{y}_{calib}$ denotes the k-space of calibration data and $\mathcal{R}$ represents the operations of readout concatenation with CAIPI shifts, as described in Eq.~\ref{eq:2}. The estimated kernels $kernel_{\theta}$ are then used to synthesize (interpolate) the missing k-space data in the low-frequency region, which can be expressed as:
\begin{equation}
\label{eq:predict Low-frequency}
\mathbf{y'} = \text{LFE}_{s}(kernel_{\theta}, \mathbf{y})
\end{equation}
where $\mathbf{y'}$ represents the updated k-space with both the LFE estimated and original data, and $s$ denotes the size of the LFE, specifically the side length of the k-space region in which missing data are synthesized. In this study, $s$ is fixed at 8 and its choice is validated in the ablation study in section~\ref{ablationstudy}. Finally, the sampling mask in $\mathcal{A}$ for reverse diffusion sampling as described in section~\ref{reverse_sampling} will also be updated to preserve both the LFE estimated and original data.

In summary, our reconstruction algorithm can be described with the pseudocode below.
\begin{algorithm}[H]  
\caption{SMS Reverse Diffusion Sampling} 
\begin{algorithmic}[1]  
    \Require calibration measurement $\mathbf{y}_{calib}$, SMS measurement $\mathbf{y}$, forward model $\mathcal{A}$, reorder operator $\mathcal{R}$, LFE size $s$, diffusion hyperparameters $\eta$ and $\lambda$.

    \State $kernel_{\theta} \xleftarrow{\text{fitting}} \mathcal{R}(\mathbf{y}_{calib})$
    \State $\mathbf{y'} = \text{LFE}_{s}(kernel_{\theta}, \mathbf{y})$
    
    \State $\mathbf{x}^{ms}_{T} \sim \mathcal{CN}(0,\mathbf{I})$
    \For{$t = T$ to $0$}  
        \State $\mathbf{x}^{ms}_{0|t}=\frac{1}{{\sqrt{\overline{a}_{t}}}}({\mathbf{x}^{ms}_{t}-\sqrt{1-\overline{a}_{t}}\epsilon^{t}_{\theta}(\mathbf{x}^{ms}_{t})})$
        
        \State $\mathbf{\hat{x}}^{ms}_{0|t} = \mathbf{x}^{ms}_{0|t} - \lambda\mathcal{(AR)}^{H}(\mathcal{AR}\mathbf{x}^{ms}_{0|t} - \mathbf{y'})$
        
        \State $\mathbf{x}^{ms}_{t-1} = \sqrt{\overline{a}_{t-1}}\mathbf{\hat{x}}^{ms}_{0|t} + \sqrt{1-\overline{a}_{t-1}}(\sqrt{1-\eta^2}\epsilon^{t}_{\theta}(\mathbf{x}^{ms}_{t}) + \eta \mathbf{z}), \mathbf{z} \sim \mathcal{CN}(0,\mathbf{I})$
    \EndFor \\
    \Return $\mathbf{x}^{ms}_{0}$
\end{algorithmic}
\end{algorithm}

\section{EXPERIMENTS}
\subsection{Datasets}
First, three raw k-space datasets were used with retrospective SMS acceleration to evaluate our method:

\begin{enumerate}
    \item The public fastMRI brain dataset\citep{zbontar2018fastmri}. In brief, this dataset includes brain anatomical imaging data acquired on 1.5T and 3T magnets. The official training set was used to train our model (see section \ref{Implementation} for details). In the training set, the majority were T2-weighted scans (2678 volumes), with the rest being T1-weighted (1447 volumes) and fluid attenuated inversion recovery (FLAIR) scans (344 volumes). We randomly selected 16 scans of T2-weighted contrast from the official validation set for method evaluation. Note that for data de-identification, this dataset does not contain any slices more than 5mm below the orbital rim\citep{zbontar2018fastmri}.
    
    \item An in-house clinical dataset (Longgang). It included T1-weighted and T2-FLAIR images from 8 subjects with white matter lesions, acquired using a 3T Siemens scanner equipped with a 20-channel head coil. The common scan parameters were FOV = 220 mm and slice thickness/gap = 5/1.5 mm. For T1-weighted scans, TR/TE = 250/2.49 ms, flip angle (FA) = 70{\textdegree}, and matrix size = 320$\times$288. For FLAIR, TR/TE/TI = 8000/84/2370 ms, refocusing FA = 150{\textdegree}, and matrix size = 320$\times$224.
    
    \item An in-house T2-FSE dataset (Huaxin) from 4 healthy volunteers using a 3T GE scanner equipped with a 21-channel head coil. This dataset has whole brain coverage including the cerebellum and brainstem. The scan parameters were TR/TE = 4784/100 ms, refocusing FA = 111{\textdegree}, matrix size = 256$\times$256, FOV = 220 mm, slice thickness/gap = 3/0.3 mm, and 48 slices.
\end{enumerate}

For retrospective acceleration, we experimented with MB factors of 3 and 4, and in-plane acceleration R factors of 2 and 3, resulting in four acceleration combinations: MB3R2, MB3R3, MB4R2, and MB4R3. This notation, MB$\times$Ry, will be used in the following sections to denote the specific acceleration settings.

Three prospectively SMS-accelerated datasets were used to further validate our methods:

\begin{enumerate}
    \item Prospectively SMS-accelerated T2-FSE data from one healthy subject at MB3R3 and MB4R2 using a 3T Siemens scanner equipped with a 64-channel head coil. The scan parameters were TR/TE = 6000/100 ms, refocusing FA = 150{\textdegree}, matrix size = 320$\times$320, FOV = 220 mm, slice thickness/gap = 2/0.4 mm, and 30 slices. The vendor-provided SMS sequence was used, which acquired separate coil calibration data for each slice (see Fig.~\ref{fig:LFE_illu} for visualization). Additionally, fully sampled T2-FSE data were acquired as a reference for desired image quality.
    
    \item Prospectively SMS-accelerated gradient echo single-shot EPI data from six healthy subjects at MB4R2, MB2R2, and MB1R1 (i.e., no acceleration), using a 3T GE scanner equipped with a 21-channel head coil. The vendor-provided SMS sequence was used and common scan parameters were matrix size = 128$\times$128, FOV = 220 mm, slice thickness/gap = 3/0.3 mm, and 48 slices. For MB1R1, TR/TE = 4480/30 ms and FA = 90{\textdegree}; for MB2R2, TR/TE = 2240/30 ms and FA = 84{\textdegree}; for MB4R2, TR/TE = 1120/30 ms and FA = 71{\textdegree}.

    \item Prospectively SMS-accelerated  functional MRI (fMRI) data using gradient echo single-shot EPI sequence. The data were acquired from a single healthy subject performing a visual stimulation task using a 3T Siemens scanner equipped with a 32-channel head coil. The acquisition was accelerated at MB4R2, with the following scan parameters: TR/TE = 1000/30 ms, FA = 62{\textdegree}, matrix size = 128$\times$128, FOV = 220 mm, slice thickness/gap = 3/0 mm, and 48 slices. During the scan, the subject alternated between 20-second checkerboard stimulation and 20-second rest periods in a block design paradigm, totaling 300 seconds. 

\end{enumerate}

For all in-house datasets, the subjects provided written informed consent, in accordance with the approval from the Institutional Review Board of Shenzhen Technology University (ref no. SZTULL012021005).

\subsection{Method Comparison and Implementation Details} \label{Implementation}

We compared our method with two widely used k-space interpolation techniques RO-GRAPPA\citep{moeller2014ro} and SPSG\citep{cauley2014interslice}, two traditional iterative methods L1-wavelet SENSE\citep{uecker2014espirit} and ROCK\citep{demirel2021improved}, and one representative supervised deep learning method VarNet\citep{sriram2020e2evarnet}. 

For RO-GRAPPA, SPSG, L1-wavelet, and ROCK, we used the central 64$\times$64 k-space region as the calibration signal. For L1-wavelet and ROCK, we estimated coil sensitivity maps using the ESPIRiT\citep{uecker2014espirit} method from the central 30$\times$30 k-space region. The choices of k-space calibration sizes were based on common practices in prior works\citep{pytorchgrappa,uecker2014espirit} and empirical adjustments on collected datasets.

For deep learning methods, i.e., VarNet and ROGER, the models were trained on the official fastMRI brain training set. We removed the last four noisy slices from each volume, resulting in approximately 52k slice images for training. The models were then used to infer on all four FSE datasets without fine-tuning. For SMS EPI datasets, light fine-tuning was performed as detailed in section \ref{GE_EPI_RESULTS}.

For VarNet, paired SMS data and fully sampled data were created under the ROC framework and used to train VarNet weights for 50 epochs with the official settings\citep{zbontar2018fastmri}. For ROGER, multi-coil images were coil-combined using ESPIRiT and then used to train the diffusion generative model. The complex-valued images were split into real and imaginary components, each serving as a separate channel. The hyperparameter $\eta$ in Eq.~\ref{eq:ddpm reverse2} was set to 1, $\lambda$ in Eq.~\ref{eq:13} was set to 2, and the LFE size $s$ in Eq.~\ref{eq:predict Low-frequency} was set to 8. We adopted the UNet network with multi-resolution attention as the DDPM architecture, similar to previous work\citep{luo2023bayesian,DDS,huang2024noise}, and used the implementation from the Ablated Diffusion Model (ADM) project\citep{dhariwal2021diffusion,nichol2021improved}. Training was conducted with forward/reverse step of 1000, learning rate of $1\times10^{-4}$, batch size of 8, $2\times10^{5}$ iterations, and the Adam optimizer.

For the GE SMS EPI data, to address the geometric distortion issue\citep{Polimeni2016fleet,lyu2018improved,fMRI7TeslaReview}, MB1R1 data were used first to train GRAPPA kernels to reconstruct MB2R2 images, which were then used to generate coil sensitivity maps for MB4R2. This provides a better geometry match between the coil sensitivity maps and the SMS data due to the same in-plane acceleration factors. For the Siemens SMS EPI data, coil sensitivity calibration was conducted using single-band 2-shot EPI data with VC-SAKE\citep{lyu2018robust, liu2019pec} phase correction.

\subsection{Performance Evaluation}
For retrospective accelerated data, peak signal-to-noise ratio (PSNR) and the structural similarity index (SSIM) were used to measure the reconstruction performance. Larger PSNR and SSIM values indicate better reconstruction. In addition, we used paired t-tests to assess the statistical significance of PSNR and SSIM differences between ROGER and other methods.

For prospective acceleration, no perfect ground truth is available due to potential subject motion, different geometry distortion and slightly different contrasts. While it is possible to use reference-free image quality metrics\citep{reference-free,kustner2018machine,van2024non}, we adopted the conventional radiologist scoring approach for simplicity. Two board-certified radiologists (with 10 and 15 years of experience, respectively) independently assessed images generated by different methods. The evaluation covered five key categories: tissue contrast, sharpness, signal-to-noise ratio (SNR), artifact reduction, and overall image quality. Each category was scored on a scale from 1 to 5, where 1 indicates poor quality and 5 represents excellent quality. The scores were averaged across slices and radiologists, and standard deviations were computed. We also calculated the tSNR maps for the EPI images. fMRI analysis was performed on the Siemens EPI data using SPM12 with standard motion correction, smoothing (full width at half maximum = $2\times2\times3$ mm), and linear regression, and the activation maps were plotted with t-score threshold 3.2 and p-value threshold 0.001. 

\section{RESULTS}
This section presents both quantitative and qualitative reconstruction results. The evaluation begins with the fastMRI and Longgang datasets, followed by an assessment of performance on unseen anatomical regions using the Huaxin dataset. A detailed analysis of the prospectively SMS-accelerated FSE and EPI datasets is then provided. Additionally, the impact of EPI fine-tuning sample size and the effectiveness of the LFE module are analyzed.

\subsection{Evaluation on Public and In-House Clinical Datasets}
Table.~\ref{tabel:sota table} presents the quantitative reconstruction results for retrospective SMS acceleration on the fastMRI and Longgang datasets. Our method ROGER achieved the highest scores at all acceleration factors for both datasets, and the differences in PSNR and SSIM between ROGER and all other methods were statistically significant ($p<0.01$), except for the SSIM comparison with VarNet at MB3R2.

\begin{figure*}[h]
\centering
\includegraphics[width=1.0\linewidth]{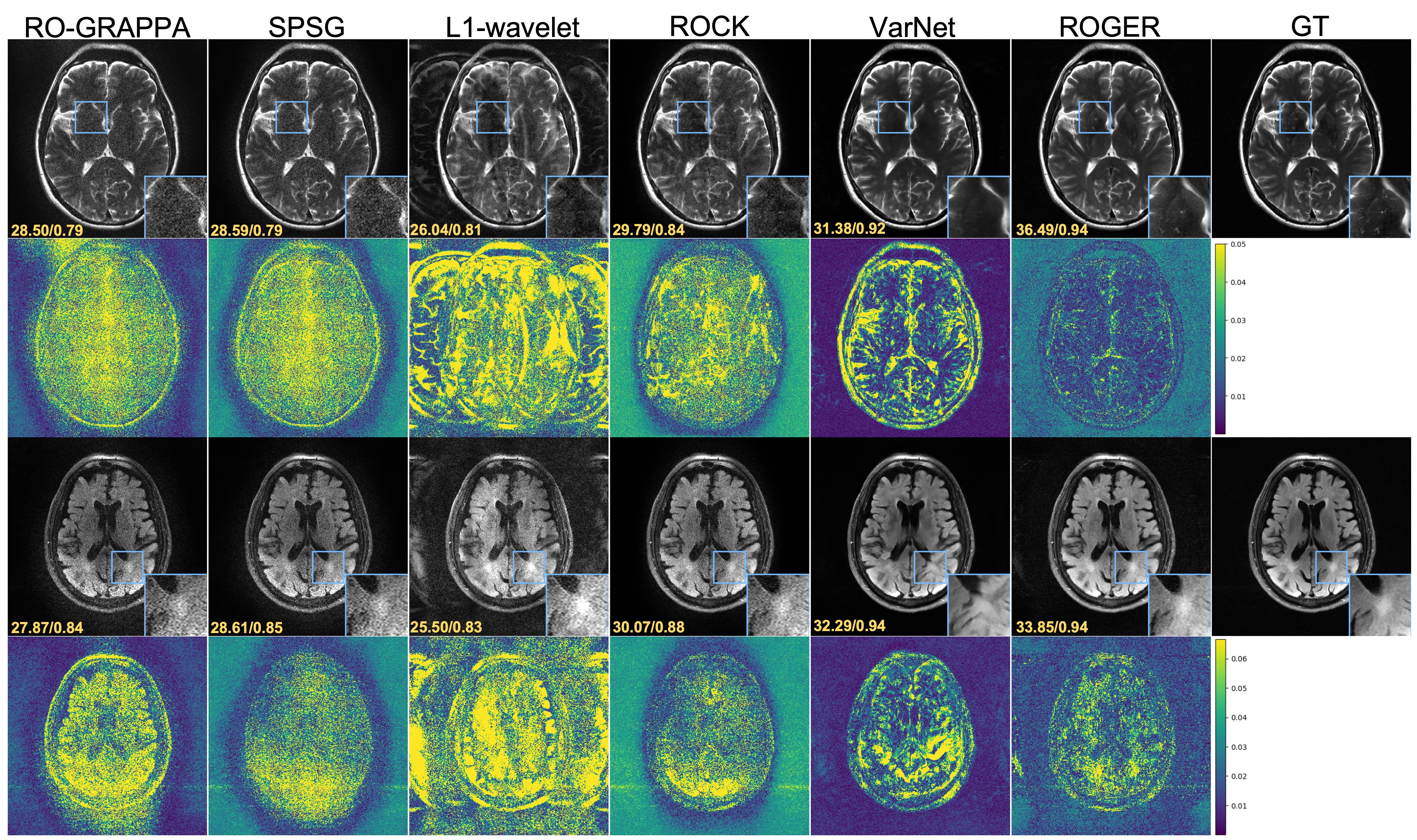}
\caption{Reconstructed images of retrospective SMS acceleration at MB4R2 in comparison with the ground truth (GT). The first and third rows are reconstructed images from the fastMRI and Longgang datasets, respectively, and their corresponding error maps are shown in the second/fourth rows. Yellow numbers represent PSNR/SSIM scores. Full visualizations across all slices/subjects are available in the supplementary materials.}
\label{fig:Retrospective1}
\end{figure*}

Representative reconstruction results are presented in Fig.~\ref{fig:Retrospective1}. GRAPPA and SPSG exhibited noticeable reconstruction noise, while L1-wavelet SENSE and ROCK showed aliasing artifacts. The results of VarNet were satisfactory, with a higher signal-to-noise ratio and fewer aliasing artifacts than traditional methods. However, VarNet failed to reconstruct many fine details and erroneously presents some white matter lesions as fake structures (see the enlarged view in Fig.~\ref{fig:Retrospective1}). In contrast, our method demonstrated the highest reconstruction quality, clearly revealing anatomical structures and small white matter lesions.

\begin{table*}[t]
\caption{Quantitative evaluation of SMS reconstruction (retrospective acceleration). The best results are marked in \textbf{Bold} and second-best \underline{underlined}.}
\label{tabel:sota table}
\begin{center}
\resizebox{1\textwidth}{!}{
\begin{tabular}{c|c|cccc|cccc}
\hline
\hline
\multirow{3}{*}{Acceleration}& \multirow{3}{*}{Method} & \multicolumn{4}{c|}{fastMRI} & \multicolumn{4}{c}{Longgang} \\
\cline{3-10}
& &  \multicolumn{2}{c}{R2} & \multicolumn{2}{c|}{R3} & \multicolumn{2}{c}{R2}& \multicolumn{2}{c}{R3} \\
\cline{3-10}
& & PSNR & SSIM & PSNR & SSIM & PSNR & SSIM & PSNR & SSIM \\
\hline
\multirow{6}{*}{MB3}
& RO-GRAPPA & 28.84±3.96 & 0.84±0.08 & 22.27±2.33 & 0.71±0.08 & 34.34±3.48 & 0.91±0.04 & 25.96±3.70 & 0.76±0.08 \\
& SPSG & 29.26±3.69 & 0.85±0.07 & 21.96±2.22 & 0.71±0.08 & 34.53±3.41 & 0.91±0.04 & 25.69±3.58 & 0.76±0.08\\ 

& L1-wavelet & 25.49±2.06 & 0.85±0.05 & 22.59±1.88 & 0.80±0.06 & 28.83±2.51 & 0.89±0.04 & 24.29±2.24 & 0.84±0.05 \\

& ROCK & 30.68±2.68 & \underline{0.89±0.04} & 25.68±2.34 & 0.83±0.05 & \underline{34.91±3.22} & \underline{0.92±0.03} & 27.92±2.67 & 0.87±0.05 \\

& VarNet & \underline{34.18±1.55} & \textbf{0.96±0.01} & \underline{31.86±1.40} & \underline{0.94±0.01} & 33.46±1.53 & \textbf{0.95±0.02} & \underline{31.49±1.60} & \underline{0.93±0.02} \\

& ROGER & \textbf{37.10±2.29} & \textbf{0.96±0.02} & \textbf{34.02±2.14} & \textbf{0.95±0.02} & \textbf{37.25±2.87} & \textbf{0.95±0.02} & \textbf{34.78±2.43} & \textbf{0.94±0.03} \\
\hline

\multirow{6}{*}{MB4}

& RO-GRAPPA & 24.76±3.71 & 0.76±0.10 & 20.71±1.78 & 0.67±0.08 & 30.77±3.52 & 0.86±0.05 & 22.60±3.06 & 0.69±0.08 \\

& SPSG & 25.05±3.68 & 0.77±0.10 & 20.51±1.68 & 0.67±0.08 & 31.04±3.46 & 0.86±0.05 & 22.22±2.93 & 0.68±0.08\\

& L1-wavelet & 23.42±1.92 & 0.81±0.06 & 21.46±1.65 & 0.77±0.06 & 25.60±2.16 & 0.86±0.05 & 22.80±2.02 & 0.80±0.06  \\

& ROCK & 27.23±2.58 & \underline{0.85±0.05} & 23.92±1.93 & \underline{0.80±0.05} & 30.97±2.67 & 0.89±0.04 & 25.46±2.22 & 0.83±0.05 \\

& VarNet & \underline{32.16±1.67} & \textbf{0.95±0.02} & \underline{30.20±1.46} & \textbf{0.93±0.01} & \underline{31.83±1.74} & \underline{0.93±0.02} & \underline{30.48±1.62} & \underline{0.91±0.03} \\

& ROGER & \textbf{35.18±2.51} & \textbf{0.95±0.02} & \textbf{31.32±2.16} & \textbf{0.93±0.02} & \textbf{36.13±2.63} & \textbf{0.95±0.02} & \textbf{32.21±2.16} & \textbf{0.92±0.03} \\

\hline
\hline
\end{tabular}}
\end{center}
\end{table*}

\subsection{Evaluation on Unseen Brain Regions}
\begin{figure*}[ht]
\centering
\includegraphics[width=1.0\linewidth]{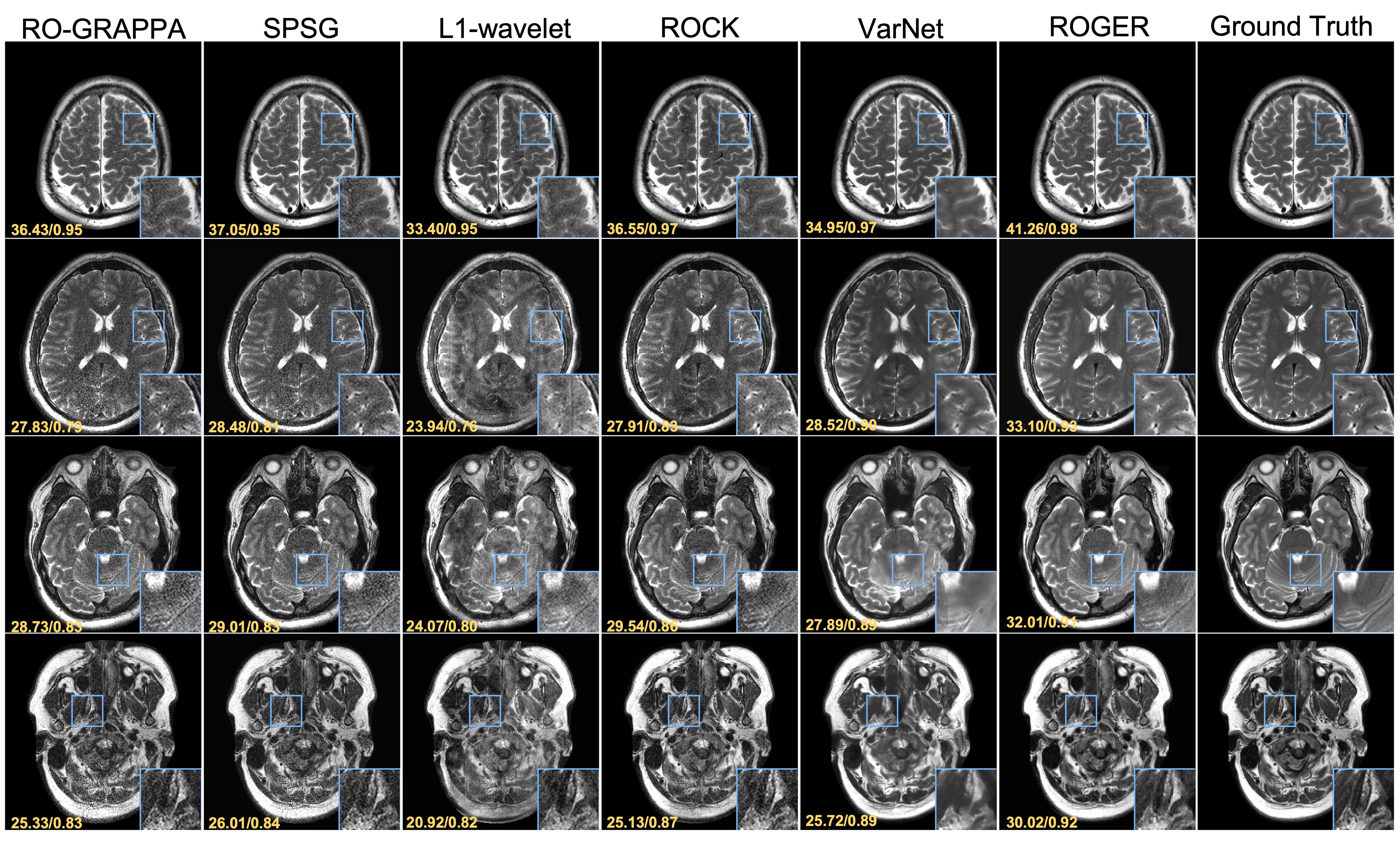}
\caption{Reconstructed images of retrospective SMS acceleration at MB4R2 on the Huaxin dataset with full brain coverage. Yellow numbers represent PSNR/SSIM scores. Four representative slices are presented with ground truth (GT). Our method ROGER remained robust on the brainstem area which was not present in the training set (fastMRI). Full visualizations across all slices/subjects are available in the supplementary materials.}
\label{fig:Retrospective2}
\end{figure*}

\begin{figure*}[h]
\centering
\includegraphics[width=0.8\linewidth]{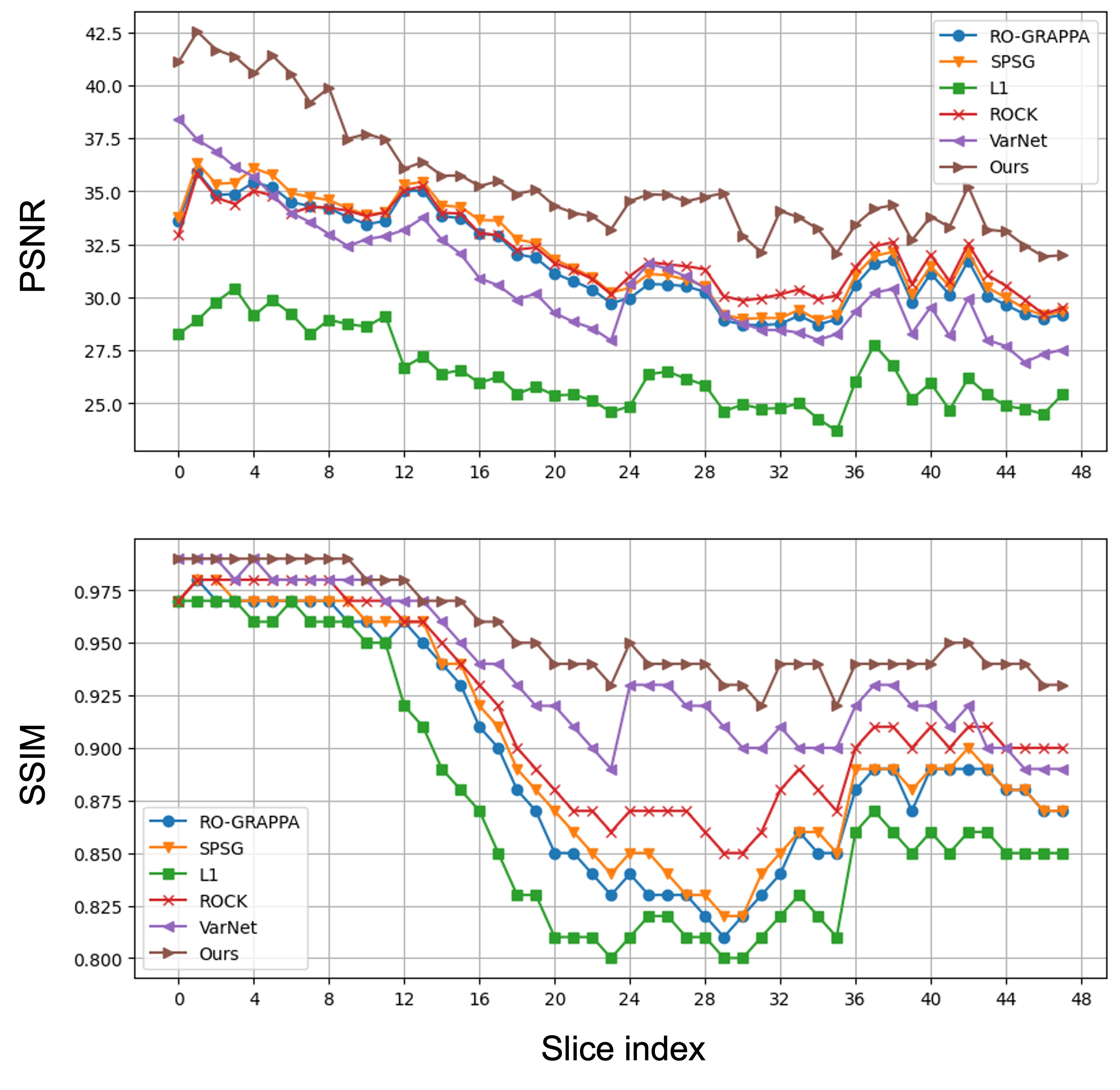}
\caption{Slice-wise PSNR and SSIM analysis for different methods at MB4R2 on one subject of the Huaxin dataset.}
\label{fig:slice_analysis}
\end{figure*}

Since the fastMRI training data does not contain slices more than 5mm below the orbital rim\citep{zbontar2018fastmri}, we used the retrospectively accelerated Huaxin dataset, covering the whole brain, including the cerebellum and brainstem, to evaluate our method on unseen anatomical regions. As indicated by the mean PSNR/SSIM values in Table~\ref{tabel:huaxin table} and the visualization in Fig.~\ref{fig:Retrospective2} at MB4R2 acceleration, our algorithm achieved the best results among all methods again, despite not being trained on the inferior brain regions. Further analysis of Table~\ref{tabel:huaxin table} revealed that ROGER's PSNR and SSIM advantages were statistically significant ($p<0.01$) compared to all other methods. Slice-level analysis presented in Fig.~\ref{fig:slice_analysis} revealed that our algorithm maintained superior performance on all slices with stable PSNR/SSIM advantages. 

\begin{table*}[t]
\caption{Quantitative evaluation of SMS reconstruction on the Huaxin datasets with whole brain coverage. The best results are marked in \textbf{Bold} and second-best \underline{underlined}.}
\label{tabel:huaxin table}
\resizebox{1.0\textwidth}{!}{
\begin{tabular}{c|cccc}
\hline
\hline
\multicolumn{1}{c|}{Acceleration} & \multicolumn{2}{c}{MB3R2} & \multicolumn{2}{c}{MB4R2} \\
\hline
Metrics & PSNR & SSIM & PSNR & SSIM\\
\hline
RO-GRAPPA  & 34.30±3.19 & 0.92±0.04 & 31.37±2.73 & 0.89±0.05 \\
SPSG       & 34.61±3.23 & 0.93±0.04 & 31.80±2.80 & 0.89±0.05 \\ 
L1-wavelet         & 29.13±2.76 & 0.90±0.05 & 26.33±2.22 & 0.87±0.06 \\
ROCK       & \underline{35.05±3.00} & 0.94±0.03 & \underline{32.07±2.62} & 0.91±0.04 \\
VarNet & 32.30±2.57 & \underline{0.95±0.02} & 30.48±2.61 & \underline{0.93±0.03} \\
ROGER      & \textbf{36.49±3.48} & \textbf{0.96±0.02} & \textbf{35.37±3.12} & \textbf{0.95±0.02} \\
\hline
\hline
\end{tabular}}
\end{table*}

\subsection{Evaluation on Prospectively Accelerated SMS FSE}
\begin{figure*}[h]
\centering
\includegraphics[width=1.0\linewidth]{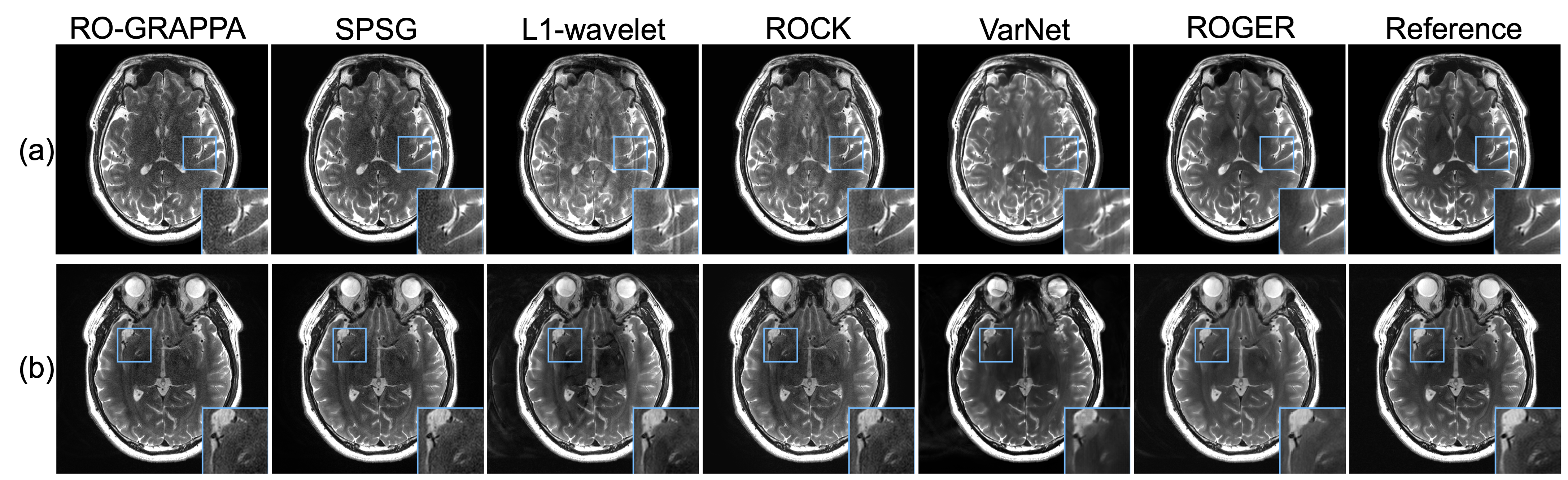}
\caption{(a) Reconstruction of prospectively SMS-accelerated FSE at MB3R3. (b) Reconstruction of prospectively SMS-accelerated FSE data at MB4R2. The reference images were acquired by a fully sampled scan. Full visualizations across all slices/subjects are available in the supplementary materials.}
\label{fig:simens_MB3R3}
\end{figure*}

The results of prospective SMS FSE acceleration agree with the findings with retrospective acceleration. As presented in Fig.~\ref{fig:simens_MB3R3}, at MB3R3 and MB4R2 acceleration, RO-GRAPPA and SPSG had similar results, both exhibiting heavy noise amplification. L1-wavelet SENSE and ROCK showed considerable aliasing artifacts. 

Directly applying the fastMRI trained VarNet resulted in excessive aliasing artifacts in prospectively SMS-accelerated data (not shown in Fig.~\ref{fig:simens_MB3R3}, available in the supplementary materials as "VarNet-Plain"). This issue likely arises because, during training, the calibration signal and SMS data had consistent image phases. However, in real-world SMS-accelerated data, the phases of the calibration signal and the SMS data are rarely perfectly aligned, leading to poor generalization of VarNet. As a workaround, we used the reconstruction results of RO-GRAPPA as calibration for VarNet, enabling it to perform reconstruction with reasonable results. Still, such improved VarNet suffered from residual artifacts and blurring on small structures as revealed by the enlarged views in Fig.~\ref{fig:simens_MB3R3}. In contrast, our method produced high-quality results that closely resembled the reference structure. This was achieved without any modifications to the fastMRI trained model or inference procedure.

The assessments by two radiologists are summarized in Table \ref{tab:qualitative_blind}. Across both MB3R3 and MB4R2 settings, ROGER consistently achieved the highest scores among all compared methods across all categories, including tissue contrast, sharpness, SNR, artifact reduction, and overall image quality.

\subsection{Evaluation on Prospectively Accelerated SMS EPI}\label{GE_EPI_RESULTS}
Because EPI has very different tissue contrast, image phase distribution, geometry distortion, and matrix sizes to routine anatomical images, we randomly selected 3 out of the 6 subjects from the GE EPI dataset to fine-tune the VarNet and ROGER model from their respective fastMRI weights. From each subject, only one frame of the MB1R1 images and one frame of the GRAPPA reconstructed MB2R2 images were used for fine-tuning after discarding the top and bottom three slices. Thus, in total, only 252 slice images were used for fine-tuning. We used a learning rate of $1\times10^{-4}$, batch size of 8, iterations of $4\times10^{4}$, and Adam optimizer. 

\begin{figure*}[ht]
\centering
\includegraphics[width=1.0\linewidth]{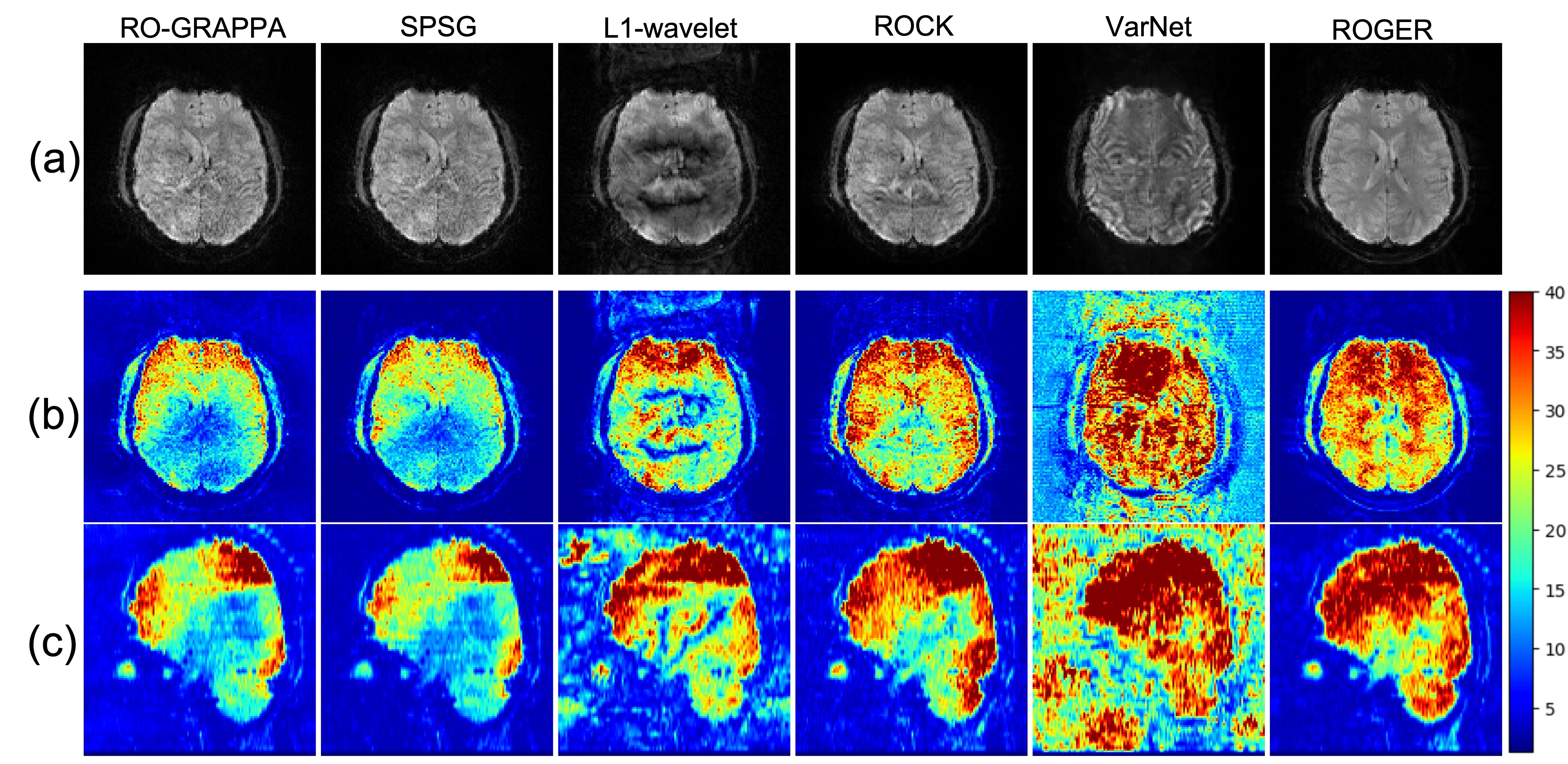}
\caption{(a) Reconstruction of prospectively SMS-accelerated EPI, acquired on a 3T GE scanner with a 21 channel coil at MB4R2 acceleration. (b) The corresponding tSNR maps computed from 50 frames. (c) The sagittal view of tSNR maps. Full visualizations across all slices/subjects are available in the supplementary materials.}
\label{fig:EPI_MB4R2}
\end{figure*}

\begin{figure*}[h]
\centering
\includegraphics[width=1.0\linewidth]{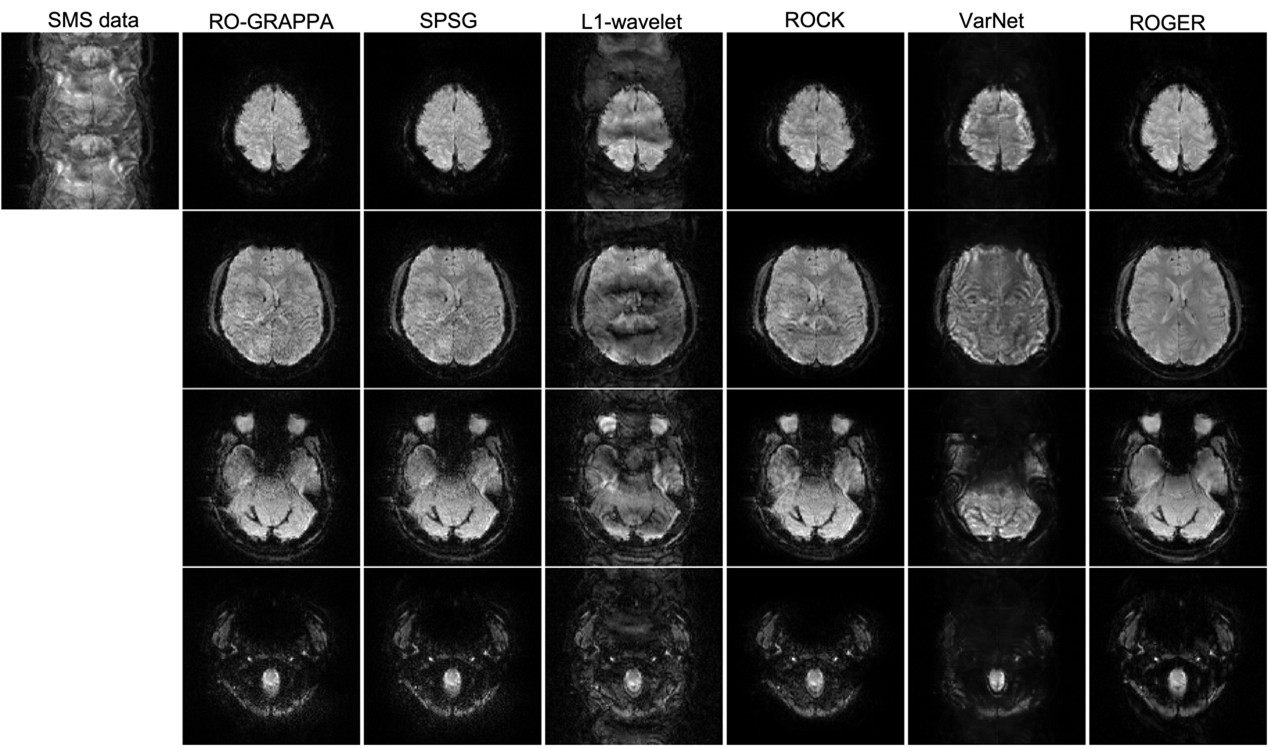}
\caption{Further demonstration of prospectively SMS-accelerated EPI reconstruction. The inverse Fourier–transformed SMS inputs are shown together with the four reconstructed slices for each method. The proposed method provides reliable reconstruction across all slices.}
\label{fig:SMS_EPI_1To4}
\end{figure*}

The MB4R2 EPI data (50 frames) acquired from the remaining three subjects on the GE scanner were used as the test set. As shown in Fig.~\ref{fig:EPI_MB4R2}, ROGER resulted in remarkably higher SNR and fewer artifacts than other methods. Moreover, our reconstruction led to consistently high tSNR across all brain regions. Fig.~\ref{fig:SMS_EPI_1To4} provides additional visualization of prospectively SMS-accelerated EPI reconstruction. By presenting the inverse Fourier–transformed SMS inputs together with the reconstructed slices, this figure illustrates the slice separation process and underscores the reliable performance of the proposed method across different slice locations.

\begin{table*}[!t]
    \caption{Two radiologists qualitatively assess the diagnostic quality of images for categories of contrast, sharpness, SNR, artifacts and overall image quality on a 1 to 5 scale (5=excellent). }
    \centering
    \label{tab:qualitative_blind}    
    \resizebox{1.0\textwidth}{!}{
    \begin{tabular}{ccccccc}
        \hline
        \hline
            Acceleration & Methods & Tissue Contrast & Sharpness & SNR & Artifacts & Overall \\
        \hline
          \multirow{6}*{MB3R3} 
& RO-GRAPPA   & \underline{3.05±0.97} & \underline{3.55±0.50} & 3.05±0.97 & \underline{3.60±0.49} & \underline{3.25±0.83} \\
& SPSG        & \underline{3.05±0.97} & \underline{3.55±0.50} & \underline{3.10±0.89} & 3.45±0.80 & \underline{3.25±0.83} \\
& L1-wavelet  & 2.70±0.56 & 2.35±0.48 & 2.75±0.43 & 2.40±0.58 & 2.40±0.49 \\
& ROCK        & 3.15±0.48 & 2.75±0.70 & 3.10±0.44 & 2.95±0.86 & 2.90±0.70 \\
& VarNet      & 2.75±0.54 & 2.75±0.54 & 2.75±0.54 & 2.70±0.56 & 2.75±0.54 \\
& ROGER       & \textbf{4.50±0.50} & \textbf{4.50±0.50} & \textbf{4.50±0.50} & \textbf{4.50±0.50} & \textbf{4.50±0.50} \\
        \hline
            \multirow{6}*{MB4R2} 
& RO-GRAPPA   & 2.65±0.65 & \underline{3.10±0.70} & 2.55±0.74 & 3.00±0.77 & 2.65±0.65 \\
& SPSG        & 2.60±0.66 & \underline{3.10±0.70} & 2.55±0.74 & 3.00±0.77 & 2.65±0.65 \\
& L1-wavelet  & 2.10±0.83 & 2.20±0.87 & 1.95±0.92 & 1.90±1.09 & 2.00±0.89 \\
& ROCK        & 2.65±0.65 & \underline{3.10±0.70} & 2.55±0.74 & 2.85±0.79 & 2.70±0.64 \\
& VarNet      & \underline{2.75±0.83} & 2.85±0.79 & \underline{2.80±0.81} & \underline{3.10±0.94} & \underline{2.85±0.79} \\
& ROGER       & \textbf{4.50±0.50} & \textbf{4.50±0.50} & \textbf{4.50±0.50} & \textbf{4.50±0.50} & \textbf{4.50±0.50} \\
        \hline
        \hline
    \end{tabular}
            }
        \end{table*}

\subsection{Impact of Sample Sizes on EPI Fine-Tuning}
\begin{figure}[h]
\centering
\includegraphics[width=0.5\linewidth]{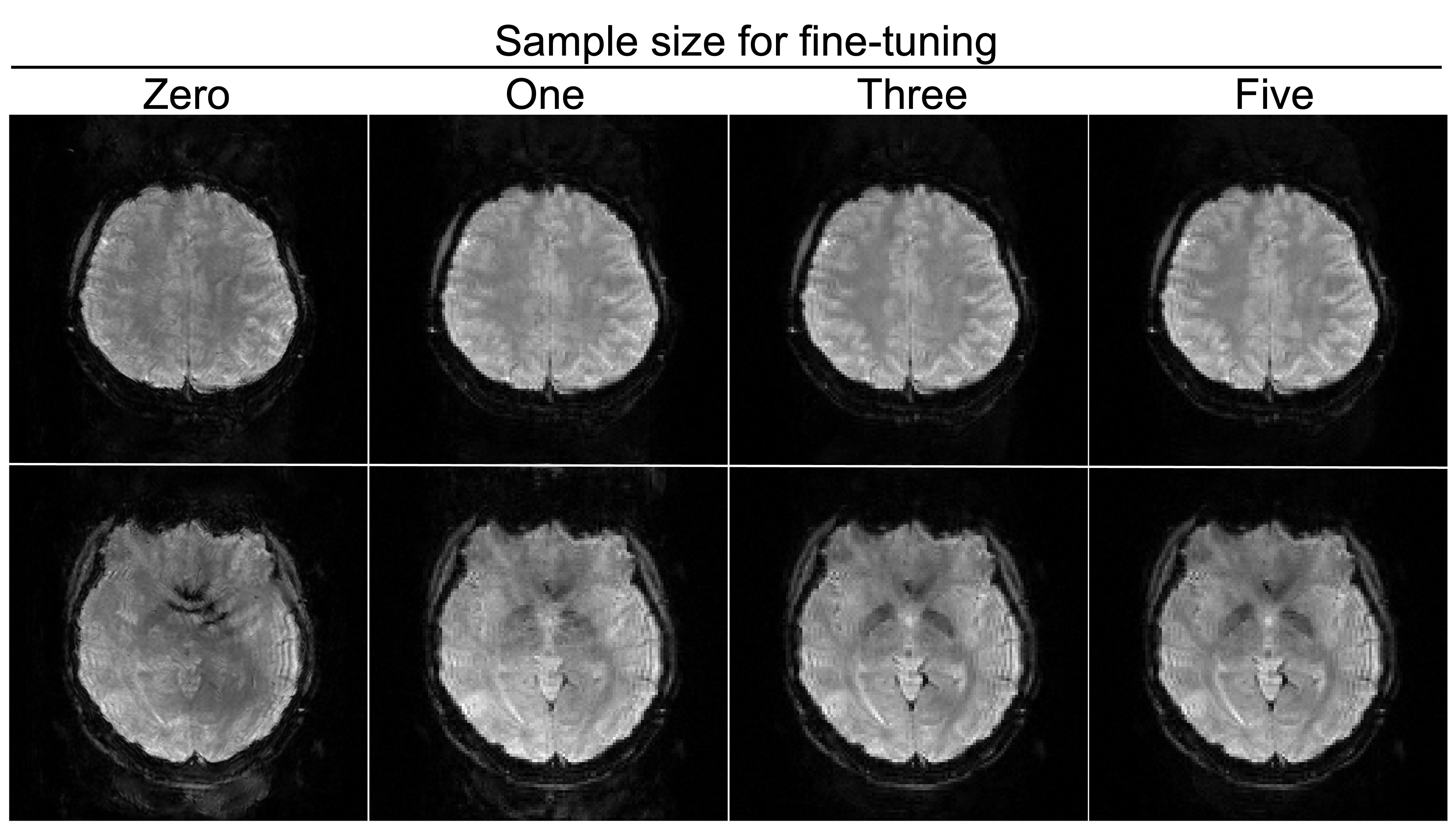}
\caption{The impact of fine-tuning sample sizes on the SMS EPI reconstruction at MB4R2. Decent reconstruction quality can be achieved with fine-tuning on three or more subjects.}
\label{fig:finetune}
\end{figure}

In addition, we studied the impact of fine-tuning sample size on SMS EPI reconstruction quality. Fig.~\ref{fig:finetune} shows the reconstruction results of one subject without fine-tuning ROGER and with fine-tuning using data from one, three, and five subjects, respectively. Without fine-tuning, the fastMRI-trained model produced noticeable artifacts due to large differences between EPI and anatomical imaging. Fine-tuning with one subject immediately improved image quality. Fine-tuning with three subjects resulted in further improvements, with no noticeable artifacts, while using five subjects yielded similar quality as three subjects. These results indicate that our method has strong generalization ability and can be applied to different datasets with minimal training resources.

\begin{figure}[h]
\centering
\includegraphics[width=1.0\linewidth]{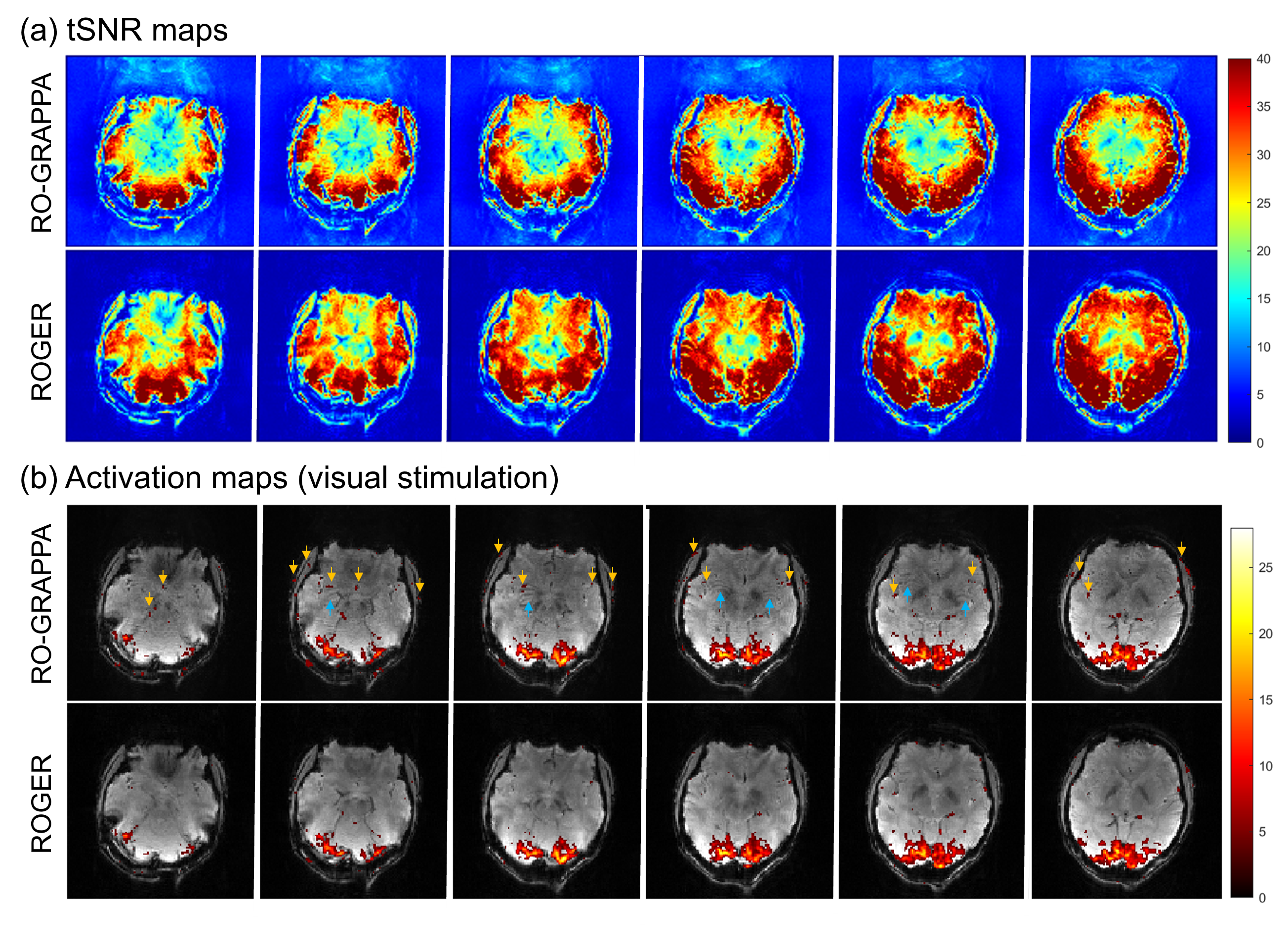}
\caption{(a) tSNR and (b) fMRI activation maps (t-score) with visual stimulation, obtained from reconstructions of prospectively SMS-accelerated EPI data (acquired on a 3T Siemens scanner at MB4R2). Yellow arrows indicate potential false positives, while blue arrows point to reconstruction artifacts. The ROGER model, despite being trained on previous GE EPI data (Fig.~\ref{fig:EPI_MB4R2}) without fine-tuning on these Siemens data, yielded improved reconstruction quality, higher tSNR, and fewer false positives than the RO-GRAPPA method.}
\label{fig:fMRI}
\end{figure}

\subsection{Out-of-Distribution EPI Reconstruction and Visual fMRI Analysis}
Fig.~\ref{fig:fMRI} presents the tSNR and fMRI activation maps for the SMS-accelerated Siemens EPI fMRI dataset (visual stimulation task). It is worth noting that the employed ROGER model was trained on previous GE EPI data (Fig.~\ref{fig:EPI_MB4R2}) without fine-tuning on this Siemens EPI dataset. Therefore, this experiment serves as an out-of-distribution test and also demonstrates the potential fMRI applications of the proposed method. As shown in Fig.~\ref{fig:fMRI}(a), the tSNR maps from ROGER exhibited higher and more uniform values across the brain, suggesting improved signal stability. In Fig.~\ref{fig:fMRI}(b), the ROGER reconstructed images displayed fewer residual artifacts compared to the RO-GRAPPA images (blue arrows), and the ROGER activation maps demonstrated fewer false positives (yellow arrows), while both methods yielded comparable true positive activations.

This observation may be explained by the fact that both methods achieved similarly high tSNR in the occipital lobe, where true positive activations are expected due to visual stimulation. However, the false positives observed in the RO-GRAPPA results were predominantly located in low tSNR regions, which are more susceptible to noise amplification and reconstruction errors. The more uniform tSNR distribution achieved by ROGER may have mitigated these effects, reducing the likelihood of spurious activations without compromising sensitivity to true signals.

\subsection{Effect of LFE Module (Ablation Study)}\label{ablationstudy}
\begin{figure*}[h]
\centering
\includegraphics[width=1.0\linewidth]{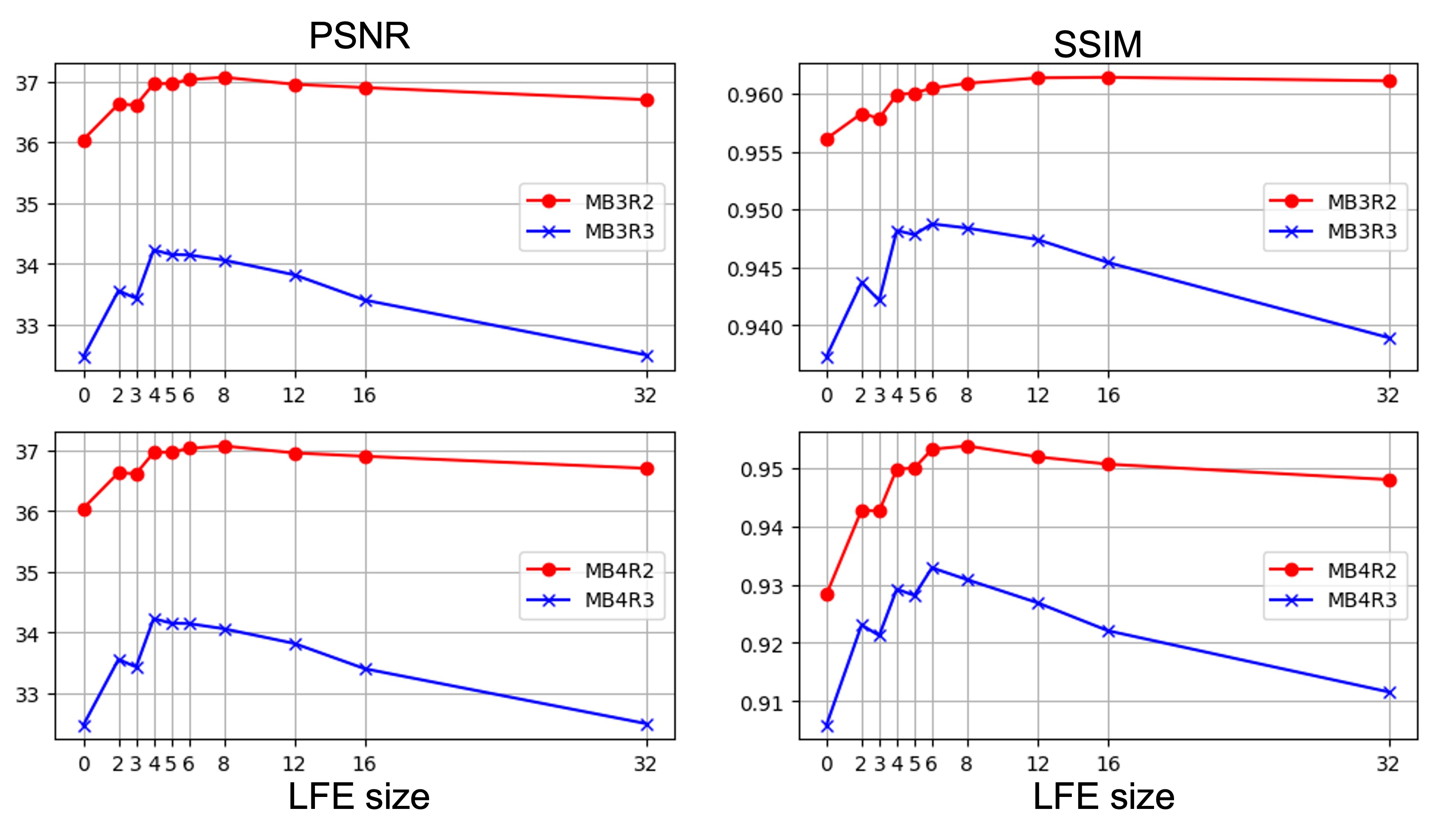}
\caption{The impact of LFE sizes on the mean PSNR/SSIM scores of the reconstructed images using the fastMRI dataset.}
\label{fig:LFE}
\end{figure*}

\begin{figure}[p]
\centering
\includegraphics[width=1.0\linewidth]{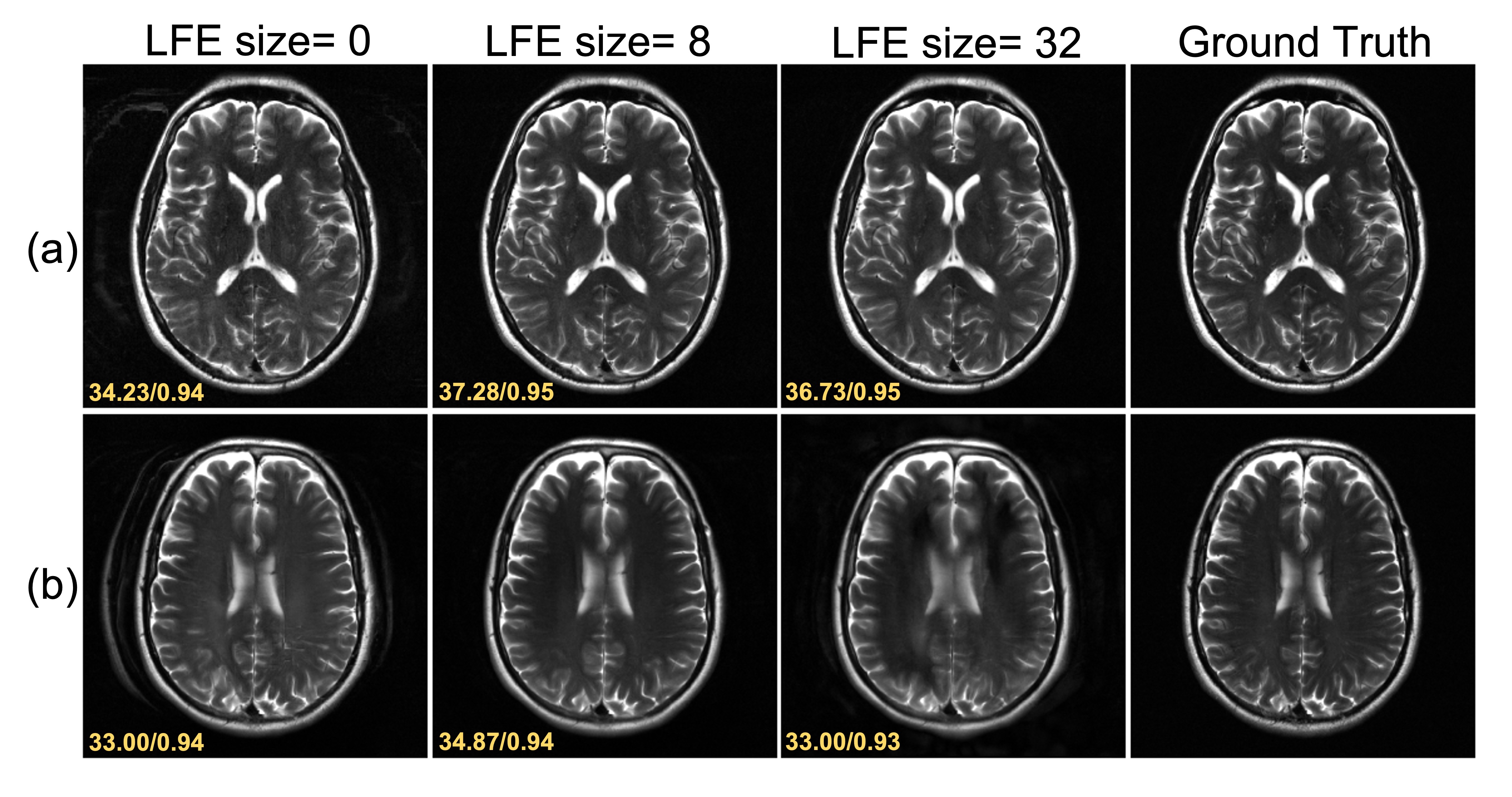}
\caption{Visualization of the impact of different LFE sizes on the reconstruction results on the fastMRI dataset (a) MB4R2 acceleration. (b) MB4R3 acceleration. Yellow numbers represent PSNR/SSIM scores. }
\label{fig:vis_LFE}
\end{figure}

To study the contribution of the proposed LFE module, Fig.~\ref{fig:LFE} records the PSNR and SSIM scores of reconstructed images with different LFE sizes (i.e., sizes of GRAPPA interpolated k-space), using the fastMRI dataset. As the LFE size increased from 0 (not using LFE) to approximately 8 (the setting in this study), our method showed marked improvement. However, excessively increasing the LFE size is not advisable, as the model performance declined with LFE sizes larger than 12, likely due to the g-factor related noise amplification\citep{grappagfactor} that GRAPPA introduced in the peripheral k-space. Such decline was more pronounced for higher acceleration (MB3R3 and MB4R3) than lower acceleration (MB3R2 and MB4R2). Nevertheless, a stable range for the LFE size existed between 4 and 12, across various acceleration factors, to consistently achieve high SNR and SSIM scores. This observation is also visualized in Fig.~\ref{fig:vis_LFE}. Note that MB4R3 represents a high acceleration scenario ($\times$12), and our method showed some loss of fine anatomical details.

\section{Discussion}
Our study introduces a novel approach SMS MRI reconstruction by integrating readout concatenation (ROC) with diffusion model-based generative priors. Our method outperforms existing techniques, achieving higher PSNR and SSIM while preserving anatomical details and reducing artifacts under various conditions. 

This method utilizes the ROC framework to apply data consistency terms, and reverses the ROC operation before applying the generative prior in each iteration. This approach ensures that data consistency terms are enforced properly while allowing the deep generative prior to be trained routinely on single-slice images without specific adjustments for SMS tasks. Without such decoupling, the prior would require complicated training on ROC images. While \citet{chung2Dsolving3D} introduced the concept of using a 2D generative prior for conventional 3D MRI reconstruction, our study addresses the added complexities of SMS MRI and uses the ROC framework to transform the SMS problem into a form that can be effectively handled by 2D priors during reconstruction.

Another key innovation in our method is the low-frequency enhancement (LFE) module. Similar concepts have been explored in previous studies\citep{sriram2020grappanet, rRAKI, PSFNet}, where known physical properties from knowledge-driven models have been integrated with deep learning techniques to enhance reconstruction reliability. As illustrated in Figs. \ref{fig:LFE} and \ref{fig:vis_LFE}, this module results in a mean PSNR improvement of 1 to 2 dB and a noticeable artifact reduction. In principle, this module is also applicable to routine in-plane accelerated FSE and EPI data. It is worth noting that, to our knowledge, few deep learning studies have successfully addressed the reconstruction of prospectively accelerated FSE or EPI data, due to the challenges associated with the lack of ACS regions, which can significantly degrade reconstruction performance\citep{SMSLORAKS}. The LFE module may provide a valuable enhancement for many deep learning methods, particularly for accelerating the widely used Cartesian FSE and EPI sequences, where embedding ACS is complicated by the T2/T2* decays of echoes. That said, the LFE module could benefit from further improvements with more advanced k-space methods\citep{SPIRIT,RAKI,rRAKI}. 

While our focus is on Cartesian SMS sequences, non-Cartesian SMS techniques\citep{bilgic2015wave,SMS-NLINV,LE2021178unetSMS,norbeck2018simultaneous,spiralsms,SMSspiral} are important due to their advantages in motion insensitivity and efficient k-space coverage. Non-Cartesian SMS sequences also allow for easier incorporation of ACS. However, reconstructing non-Cartesian data is more complex due to the need for data regridding, as well as the challenges of off-resonance effects and trajectory imperfections. Conceptually, diffusion models can be extended to non-Cartesian SMS data reconstruction, but the main challenge lies in the long reconstruction times, as each iteration requires performing NUFFT, which is computationally expensive. Furthermore, off-resonance effects and trajectory imperfections may lead to amplified artifacts. Despite these challenges, combining non-Cartesian SMS with diffusion models has the potential to enable higher spatio-temporal resolution in MRI applications.

Deep learning has profoundly impacted MRI reconstruction, yet ensuring robustness across diverse scanning conditions remains a challenge. These variations include differences in scanner hardware, imaging sequences, and scanning parameters. Through validation on multiple datasets, our method is demonstrated highly robust against most of these factors without necessitating retraining. Even in challenging EPI MRI scenarios, where existing deep learning methods may fail completely  (Fig.~\ref{fig:EPI_MB4R2}), minimal fine-tuning of ROGER on few subjects provides better results than the widely used GRAPPA method and generalizes well to another EPI dataset that has different acquisition parameters and hardware systems (Fig.~\ref{fig:fMRI}). To the best of our knowledge, this study is one of the first to demonstrate the application and benefits of diffusion models in fMRI acceleration, and we anticipate that further improvements could be achieved by training on large EPI datasets, such as HCP\citep{HCPvan2012human}. While our method provides a strong foundation, optimal performance in specialized applications might still benefit from tailored strategies\citep{huang2024zeroshot}. The potential for broader applications, including cardiac\citep{rapacchi2019simultaneous,spiralsms,ferrazzi2020autocalibrated,SMSspiral,demirel2023improved,zou2023myocardial,gaspar2023open}, knee\citep{fritz2017simultaneous}, and abdominal imaging\citep{ye2024abdominal}, suggests exciting avenues for future research.

Our study has a few limitations. First, the current implementation of ROGER relies on coil sensitivity maps estimated by ESPIRiT\citep{uecker2014espirit}, whose performance varies with calibration data quality. Second, the current setting of 1000 iterations incurs high computational loads for reconstruction (2-3 min/slice on an RTX 4090 GPU). Third, we have only used DDIM without exploring other recently proposed samplers\citep{DDS,CoreDiff,fang2025beyond,moufad2025variational}. Improving coil sensitivity estimation, accelerating reconstruction, and exploring more advanced sampling strategies remain important directions for future research.

\section{Conclusion}
In this study, we proposed a robust image reconstruction method for SMS MRI. It offers superior image reconstruction quality and high generalization ability, potentially benefiting a wide range of applications.

\setcounter{secnumdepth}{0} 
\section{CRediT Authorship Contribution Statement}
Shoujin Huang: Methodology, Software, Formal analysis, Visualization, Writing – original draft; Guanxiong Luo: Methodology, Software, Writing – review \& editing; Yunlin Zhao: Formal analysis, Visualization; Yilong Liu: Resources, Formal analysis, Funding acquisition,  Writing – review \& editing; Yuwan Wang: Data Curation, Investigation; Kexin Yang: Data curation, Investigation; Jingzhe Liu: Resources, Investigation; Hua Guo: Resources, Investigation; Min Wang: Resources, Investigation, Writing – review \& editing; Lingyan Zhang: Resources, Data curation, Project administration; Mengye Lyu: Conceptualization, Supervision, Funding acquisition, Project administration, Writing – original draft.

\section{Acknowledgments}
This work was supported in part by the National Key Research and Development Program of China (2021YFA1101700), National Natural Science Foundation of China (Grant No. 62101348, 82202096, 32301159, and 82572290), Shenzhen Higher Education Stable Support Program (Shenzhen Science and Technology Program, Grant No. 20220716111838002) and Zhejiang Provincial Natural Science Foundation of China (Z25H180012). We thank Dr. Liu Zhibo and the on-site engineers for help with data acquisition, and Dr. Wu Yin, Dr. Liu Shaojun, and Mr. Mei Lifeng for helpful discussions.

\section{Data availability}
Code, sample data, and model weights are available at \url{https://github.com/Solor-pikachu/ROGER}. The fastMRI dataset is publicly accessible, and other data can be obtained from the authors upon request.

\section{Declaration of generative AI and AI-assisted technologies in the writing process}
During the preparation of this work the author(s) used ChatGPT in order to improve readability and language. After using this tool/service, the author(s) reviewed and edited the content as needed and take(s) full responsibility for the content of the publication.

\bibliographystyle{elsarticle-harv}
\biboptions{authoryear}
\bibliography{ref}

\end{document}